\documentclass[twocolumn]{aastex631}

\usepackage{colortbl, soul}
\newcommand{\rev}[1]{#1}

\newcommand{\linelum}{$L'_\mathrm{CO}$}
\newcommand{\rhoh}{$\rho_{\mathrm{H_2}}$}
\newcommand{\tb}{$T_\mathrm{b}$}
\newcommand{\zco}{$z_\mathrm{CO}$}

\newcommand{\zopt}{$z_\mathrm{opt}$}

\newcommand{\thirteenco}{$^{13}\mathrm{CO}$}

\begin{document}

\title{COMAP Early Science: VIII. A Joint Stacking Analysis with eBOSS Quasars}

\author[0000-0002-5223-8315]{Delaney A. Dunne}
\affiliation{California Institute of Technology, 1200 E.~California Blvd., Pasadena, CA 91125, USA}

\author[0000-0002-8214-8265]{Kieran A. Cleary}
\affiliation{California Institute of Technology, 1200 E.~California Blvd., Pasadena, CA 91125, USA}

\author[0000-0001-8382-5275]{Patrick C. Breysse}
\affiliation{Center for Cosmology and Particle Physics, Department of Physics, New York University, 726 Broadway, New York, NY, 10003, USA}

\author[0000-0003-2618-6504]{Dongwoo T. Chung}
\affiliation{Canadian Institute for Theoretical Astrophysics, University of Toronto, 60 St. George Street, Toronto, ON M5S 3H8, Canada}
\affiliation{Dunlap Institute for Astronomy and Astrophysics, University of Toronto, 50 St. George Street, Toronto, ON M5S 3H4, Canada}

\author[0000-0003-3420-7766]{H\aa vard T. Ihle}
\affiliation{Institute of Theoretical Astrophysics, University of Oslo, P.O. Box 1029 Blindern, N-0315 Oslo, Norway}

\author[0000-0003-2358-9949]{J. Richard Bond}
\affiliation{Canadian Institute for Theoretical Astrophysics, University of Toronto, 60 St. George Street, Toronto, ON M5S 3H8, Canada}

\author[0000-0003-2332-5281]{Hans Kristian Eriksen}
\affiliation{Institute of Theoretical Astrophysics, University of Oslo, P.O. Box 1029 Blindern, N-0315 Oslo, Norway}

\author[0000-0002-7524-4355]{Joshua Ott Gundersen}
\affiliation{Department of Physics, University of Miami, 1320 Campo Sano Avenue, Coral Gables, FL 33146, USA}

\author[0000-0001-5211-1958]{Laura C. Keating}
\affiliation{Institute for Astronomy, University of Edinburgh, Blackford Hill, Edinburgh, EH9 3HJ, UK}

\author[0000-0002-4274-9373]{Junhan Kim}
\affiliation{California Institute of Technology, 1200 E.~California Blvd., Pasadena, CA 91125, USA}
\affiliation{\rev{Department of Physics, Korea Advanced Institute of Science and Technology (KAIST), 291 Daehak-ro, Yuseong-gu, Daejeon 34141, Republic of Korea}}

\author{Jonas Gahr Sturtzel Lunde}
\affiliation{Institute of Theoretical Astrophysics, University of Oslo, P.O. Box 1029 Blindern, N-0315 Oslo, Norway}

\author[0000-0002-8659-3729]{Norman Murray}
\affiliation{Canadian Institute for Theoretical Astrophysics, University of Toronto, 60 St. George Street, Toronto, ON M5S 3H8, Canada}

\author[0000-0002-8800-5740]{Hamsa Padmanabhan}
\affiliation{Departement de Physique Th\'eorique, Universit\'e de Gen\`{e}ve, 24 Quai Ernest-Ansermet, CH-1211 Gen\`{e}ve 4, Switzerland}

\author[0000-0001-7612-2379]{Liju Philip}
\affiliation{Jet Propulsion Laboratory, California Institute of Technology, 4800 Oak Grove Drive, Pasadena, CA 91109, USA}
\affiliation{\rev{Brookhaven National Laboratory, P.O. Box 5000, Upton, NY 11973-5000}}

\author[0000-0001-5301-1377]{Nils-Ole Stutzer}
\affiliation{Institute of Theoretical Astrophysics, University of Oslo, P.O. Box 1029 Blindern, N-0315 Oslo, Norway}

\author[0000-0002-3155-946X]{Do\u{g}a Tolgay}
\affiliation{Canadian Institute for Theoretical Astrophysics, University of Toronto, 60 St. George Street, Toronto, ON M5S 3H8, Canada}
\affiliation{Department of Physics, University of Toronto, 60 St. George Street, Toronto, ON, M5S 1A7, Canada}

\author[0000-0003-3821-7275]{Ingunn Katherine Wehus}
\affiliation{Institute of Theoretical Astrophysics, University of Oslo, P.O. Box 1029 Blindern, N-0315 Oslo, Norway}


\author[0000-0003-2358-9949]{Sarah E.~Church}
\affiliation{Kavli Institute for Particle Astrophysics and Cosmology \& Physics Department, Stanford University, Stanford, CA 94305, US}

\author{Todd Gaier}
\affiliation{Jet Propulsion Laboratory, California Institute of Technology, 4800 Oak Grove Drive, Pasadena, CA 91109, USA}

\author[0000-0001-6159-9174]{Andrew I.~Harris}
\affil{Department of Astronomy, University of Maryland, College Park, MD 20742}

\author{Richard Hobbs}
\affil{Owens Valley Radio Observatory, California Institute of Technology, Big Pine, CA 93513, USA}

\author[0000-0002-5959-1285]{James W.~Lamb}
\affil{Owens Valley Radio Observatory, California Institute of Technology, Big Pine, CA 93513, USA}

\author{Charles R.~Lawrence}
\affiliation{Jet Propulsion Laboratory, California Institute of Technology, 4800 Oak Grove Drive, Pasadena, CA 91109, USA}

\author[0000-0001-9152-961X]{Anthony C.S.~Readhead}
\affiliation{California Institute of Technology, 1200 E. California Blvd., Pasadena, CA 91125, USA}

\author{David P.~Woody}
\affil{Owens Valley Radio Observatory, California Institute of Technology, Big Pine, CA 93513, USA}

\begin{abstract}
We present a new upper limit on the cosmic molecular gas density at $z=2.4-3.4$ obtained using the first year of observations from the CO Mapping Array Project (COMAP). COMAP data cubes are stacked on the 3D positions of 243 quasars selected from the Extended Baryon Oscillation Spectroscopic Survey (eBOSS) catalog, yielding a 95\% upper limit for flux from CO(1-0) line emission of 0.129 Jy km/s. Depending on \rev{the balance of the emission between the quasar host and its environment}, this value can be interpreted as an average CO line luminosity $L'_\mathrm{CO}$ of eBOSS quasars of $\leq 1.26\times10^{11}$ K km pc$^2$ s$^{-1}$, or an average molecular gas density $\rho_\mathrm{H_2}$ in regions of the universe containing a quasar of $\leq 1.52\times10^8$ M$_\odot$ cMpc$^{-3}$. The $L'_\mathrm{CO}$ upper limit falls among CO line luminosities obtained from individually-targeted quasars in the COMAP redshift range, and the $\rho_\mathrm{H_2}$ value is comparable to upper limits obtained from other Line Intensity Mapping (LIM) surveys and their joint analyses. Further, we forecast the values obtainable with the COMAP/eBOSS stack after the full 5-year COMAP Pathfinder survey. We predict that a detection is probable with this method, depending on the CO properties of the quasar sample. Based on the achieved sensitivity, we believe that this technique of stacking LIM data on the positions of traditional galaxy or quasar catalogs is extremely promising, both as a technique for investigating large galaxy catalogs efficiently at high redshift and as a technique for bolstering the sensitivity of LIM experiments, even with a fraction of their total expected survey data.

\end{abstract}

\keywords{\href{http://astrothesaurus.org/uat/262}{CO line emission (262)}; \href{http://astrothesaurus.org/uat/336}{Cosmological evolution (336)}; \href{http://astrothesaurus.org/uat/734}{High-redshift galaxies (734)}; \href{http://astrothesaurus.org/uat/1073}{Molecular gas (1073)}; \href{http://astrothesaurus.org/uat/1319}{Quasars (1319)}; \href{http://astrothesaurus.org/uat/1338}{Radio astronomy (1338)}}

\section{Introduction} \label{sec:intro}

Spectral Line Intensity Mapping (LIM) is an emerging observational technique with the potential to enhance our understanding of the universe by constraining the global properties of galaxies over cosmic time. LIM surveys do not aim to resolve individual galaxies, but instead measure 3-dimensional fluctuations in the integrated emission from many galaxies, allowing for efficient mapping of galaxies across large cosmic volumes \citep[see][for a review]{kovetz2019_limwhitepaper}. Because LIM measures integrated emission, it is sensitive to the faintest galaxies, which are nearly impossible to detect in traditional surveys but nevertheless make up the bulk of the galaxy population at any given cosmic time. Observationally, the field is still in its early stages, with dedicated LIM instruments only beginning to publish early autocorrelation constraints (eg.~COMAP, \citealt{cleary2021_comapoverview}; the Hydrogen Epoch of Reionization Array, HERA, \citealt{hera2021_earlyscience}; and MeerKAT\rev{,} \citealt{paul2023_meerkat}).

Currently, the results being released from these first LIM surveys are primarily upper limits \citep[for COMAP,][]{chung2021_comapforecasts} and are not yet sufficient to achieve the measurements described above. This is partially a consequence of the newness of the field -- strong autocorrelation detections of a given spectral line are required to measure its cosmological fluctuations, and most LIM surveys have simply not been integrating long enough to achieve the needed sensitivities \citep[for example, COMAP projects a secure autocorrelation detection after 5 years;][]{chung2021_comapforecasts}. However, cosmic line emission may be detectable even with present-day LIM datasets, if careful signal processing techniques making use of external datasets are applied. 

In particular, averaging together intensity-map voxels (3D pixels) that are known to contain galaxies (by comparison with some other survey) in a stacking analysis has the potential to make a detection of the average CO line temperature associated with catalog galaxies, even with an intensity map insufficiently sensitive for other analyses \citep{silva2021_hetdexcomap}. Stacking, or coadding, is an established technique for improving sensitivity in traditional, targeted galaxy surveys \citep[e.g.,][]{stanley2019_quasarstack, jolly2021_almaclusterstack, romano2022_alpinestack, lujanniemeyer2022}, and has recently been extended to LIM observations as well \citep{keenan2021_copssstack}. The efficacy of this technique will depend on factors such as the number of traditional survey objects that fall into the LIM survey footprint, and the redshift accuracy in the traditional catalog \citep{chung2019_crosscorrelation, silva2021_hetdexcomap}, as well as the chosen weighting scheme \citep{sinigaglia2022_symmstacking}.

In this paper, we use LIM data from the CO Mapping Array Project \citep[COMAP;][]{cleary2021_comapoverview}, the first survey to use a purpose-built instrument \citep[the COMAP Pathfinder, described in detail in][]{lamb2021_instrument}, to impose direct constraints on the clustering-scale CO power spectrum at high redshifts. We use COMAP Season 1 data, taken during the first 13-month observing season of the project. COMAP observations cover a frequency range of 26--34\,GHz (redshifts of $z=2.4$--$3.4$) and encompass three $\sim 4$\,deg$^2$ fields. The data reduction and map-making processes for these Season 1 data are described in \cite{foss21_comapmapmaking}, while power spectrum estimation techniques and constraints are described in \cite{ihle2021_powerspectrum} and \cite{chung2021_comapforecasts}. Observations are continuing with the Pathfinder to complete the nominal 5-year survey, at the end of which a detection of the CO auto-power spectrum is forecast.  The COMAP fields were selected to overlap with the HETDEX survey of Ly-$\alpha$ emitters \citep{gebhardt2021_hetdexoverview}, allowing us to perform stacking and cross-correlation analyses of these two datasets. 

As a preliminary step, we investigate in this work the potential for the BOSS/eBOSS quasar sample \citep{2016AJ....151...44D} to provide additional constraints and inform our understanding of the relationship between active galaxies and star formation. The eBOSS catalog was developed with the intention of studying the Ly$\alpha$ forest and therefore covers a redshift range overlapping with that of COMAP -- the eBOSS DR16 catalog includes spectroscopic observations of $> 239, 000$ quasars in the range $z > 2.1$. Depending on the assumptions made, a stack of COMAP data on the positions of eBOSS quasars can be treated as a measurement of either the average CO luminosity of the quasars themselves or the average molecular gas density at quasar positions.

In addition to its extensive size, which allows for a high-sensitivity stack, the combination of COMAP with the eBOSS catalog enables the study of the CO properties of a large sample of high-redshift quasars. Quasar feedback, particularly in relation to the molecular gas traced by CO transitions, is poorly understood on a statistical scale. Current studies mainly involve individual objects, using the CO rotational ladder in the host galaxies of Active Galactic Nuclei (AGN) to study the effects of AGN on their surroundings. These are complex measurements requiring extremely CO-bright objects, and thus the pool of available galaxies for study is small and biased, especially at high redshifts. LIM measurements will be able to provide a more complete picture of the wider population of AGN and their host galaxies \citep[e.g.,][]{breysse2019_agnfeedback}. 

In this work, we will investigate the CO properties of the 243 eBOSS quasars in the COMAP fields through stacking, thus providing constraints on the molecular gas content of a large sample of high-redshift quasars for the first time. We will additionally use this analysis to explore the viabilty of stacking as a LIM technique. As such, we will forecast the sensitivity of the COMAP Pathfinder survey to CO emission from the eBOSS quasars after five years of observing. Section \ref{sec:data} introduces both the COMAP Pathfinder (\S\ref{ssec:comap_data}) and eBOSS surveys (\S\ref{ssec:eboss_data}), and Section \ref{sec:stackmethods} describes the stacking methodology we use in our analysis. \rev{Section \ref{sec:data_verification} describes verification of the stacking analysis. We present results in Section \ref{sec:results}, including forecasts of the results of stacking the full Pathfinder survey (\S\ref{ssec:forecasts}).} We discuss the implications of these results in Section \ref{sec:discussion}.

We assume a $\Lambda$CDM cosmology based broadly on nine-year WMAP \citep{Hinshaw_2013} results throughout, with $\Omega_\mathrm{m} = 0.286$, $\Omega_\Lambda = 0.714$, $\Omega_\mathrm{b} = 0.047$, and $H_0=100h\mathrm{km\ s^{-1}\ Mpc^{-1}}$ with $h=0.7$. These are the same values used in all previous COMAP works. \rev{Unless otherwise indicated}, distances are given as proper values.

\section{Data}\label{sec:data}

\subsection{COMAP}\label{ssec:comap_data}

The COMAP Season 1 data consist of 13 months of observations using the COMAP Pathfinder instrument, a single-polarization 19-feed spectrometer array fielded on a 10.4\,m telescope located at the Owens Valley Radio Observatory (OVRO) in California (the Pathfinder is described in detail in \citealt{lamb2021_instrument}). The Pathfinder instrument observes between 26 and 34\,GHz, and is therefore sensitive to the CO(1--0) rotational transition ($\nu_0 = 115.27$\,GHz) emitted in a redshift range of $z=2.4-3.4$.  These observations were towards three $\sim 4$-deg$^2$  cosmological fields (shown in Table \ref{tab:fieldlocations}).

\begin{table}[h!]
\centering
\caption{The locations of the three fields used in the COMAP Pathfinder survey.\label{tab:fieldlocations}}
\begin{tabular}{l|c|c}
\hline
\hline
Field & RA (J2000) & Dec (J2000) \\
\hline
Field 1& 01$^{\rm h}$ 41$^{\rm m}$ 44$\fs$4 & 00$\degr$ 00$\arcmin$ 00$\farcs$0 \\
Field 2& 11$^{\rm h}$ 20$^{\rm m}$ 00$\fs$0 & 52$\degr$ 30$\arcmin$ 00$\farcs$0 \\
Field 3& 15$^{\rm h}$ 04$^{\rm m}$ 00$\fs$0 & 55$\degr$ 00$\arcmin$ 00$\farcs$0 \\
\hline
\end{tabular}
\end{table}

During observations, the telescope is positioned at the leading edge of the field and performs a scanning motion while the cosmological field drifts through. This usually takes 3--10 minutes, after which the telescope is pointed to the leading edge of the field again and the procedure is repeated. This means that over time, we are scanning the fields from several different directions leading to a smoothly varying noise distribution in the final maps, with low noise in the central regions of the field, and gradually increasing levels of noise towards the edges of the fields. 

Raw time-ordered data (TOD) from the telescope are processed by applying a series of filters with the goal of removing correlated noise, standing waves, continuum foregrounds, ground contamination and other systematic effects from the TOD. This process and the COMAP scanning strategies are described in detail by \cite{foss21_comapmapmaking}. The output of this pipeline is a set of three calibrated three-dimensional intensity maps, each of angular size $\sim4$\,deg$^2$ with 31.25\,MHz spectral resolution and a 4.9--4.4\,arcmin beam FWHM. At small scales these maps are dominated by uncorrelated, Gaussian noise \citep{foss21_comapmapmaking}. The maps are somewhat elongated in the right ascension direction, giving us appreciable coverage in a 3D region with dimensions (in comoving coordinates) roughly 300\,Mpc $\times$ 200\,Mpc in directions perpendicular to the line of sight (at redshift $z = 2.9$) and about 1000\,Mpc along the line of sight, for a total volume of the order 6$\times 10^7$ Mpc${}^3$ comoving for each field. 

As described by \citet{foss21_comapmapmaking}, \citet{ihle2021_powerspectrum} and \citet{rennie2022_comapgps}, the maps are calibrated to the total power entering the telescope. About 72 \% of this power comes from the main beam, with another roughly 10 \% of the power in the near sidelobes (within about 30 arcmin). To take this scale-dependence into account \citet{ihle2021_powerspectrum} use a model of the COMAP beam to construct a beam transfer function for the power spectrum. For the purposes of our stacking analysis, since we are mostly interested in the smallest scales in the map, we simply divide the map (and the corresponding uncertainty map) by a single overall factor of 0.72 to get the right calibration at scales corresponding to the main beam. We mask any map voxel \rev{(3-D pixel)} with an integration time below 1000 seconds (50000 `hits', a stricter cut than is taken for the overall COMAP pipeline; \citealt{foss21_comapmapmaking}).

\subsection{eBOSS}\label{ssec:eboss_data}

The Baryon Oscillation Spectroscopic Survey (BOSS; \citealt{dawson2013_sdssbossreduction}) and its extension (eBOSS; \citealt{eboss_dr16}) together encompass eleven years of observations using the BOSS spectrograph at Apache Point Observatory \citep{gunn2006_sdsstelescope, smee2013_bossspectrograph}. We use the large ($\sim 240 000$-object) eBOSS spectroscopic sample of Quasi-Stellar Objects (QSOs) at $z\geq 2.4$, which was intended to enable studies of the Ly$\alpha$ forest at $z\sim 1$. These observations targeted objects that were selected based on WISE/SDSS DR13 imaging data. We use the eBOSS DR16 superset catalog \citep{sdssdr16, eboss_dr16}, released in 2020 -- the final iteration of the BOSS/eBOSS catalog. While principally composed of traditional quasars, the superset applied no quasar-selection pipeline and thus contains a small percentage (3.3\%, in the overlap with COMAP fields) of bright galaxies, broad absorption line quasars, and damped Ly$\alpha$ systems. 

\begin{figure}[ht]
    \centering
    \includegraphics[width=0.46\textwidth]{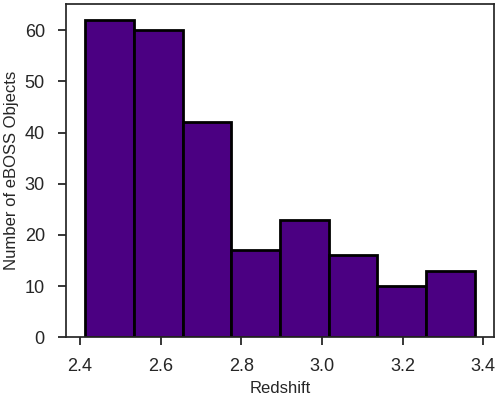}
    \caption{The redshift distribution of the eBOSS objects used for this stacking analysis.}
    \label{fig:qso_redshift_distribution}
\end{figure}

\begin{figure*}[ht]
    \centering
    \includegraphics[width=\textwidth]{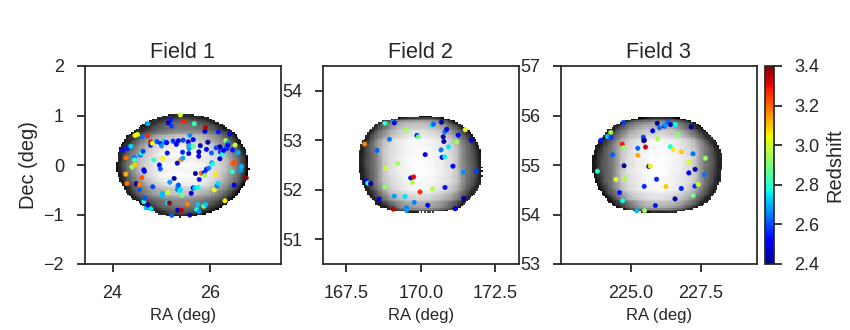}
    \caption{The distribution in space and redshift of the eBOSS objects used for this stacking analysis, plotted on the COMAP Season 1 median map RMS for each field \citep[see][]{foss21_comapmapmaking}.}
    \label{fig:qso_distribution}
\end{figure*}

Objects in the eBOSS catalog are identified and their redshifts are determined simultaneously, through the fitting of templates spanning galaxies, quasars, and stars to the eBOSS spectra \citep[stepping over redshift;][]{bolton2012_bossspectralreduction}. Spectra are optical, and at the redshifts of interest the templates to be fit typically include some combination of Ly$\alpha$, H$\alpha$ and H$\beta$, and a range of optical and NUV forbidden metal lines.  The fit with the minimum reduced-$\chi^2$ value is used, provided $\chi_r^2 > 0.01$ and there is no other similarly confident fit determination greatly offset in redshift from the best-fit value. Visual followup is used to confirm the least confident identifications, and to test the algorithmic determination\footnote{Other redshift values, in particular those based on specific emission lines, are also available. For a full list of the various methods for redshift determination, see: \url{https://data.sdss.org/datamodel/files/BOSS\_QSO/DR16Q/DR16Q\_Superset\_v3.html}}. Tests of the classification algorithm on blank-sky spectra showed a $< 2$\% false positive rate -- 
the eBOSS redshift determinations are generally considered highly trustworthy. Based on the pixel size of the detector, the uncertainty in the eBOSS redshifts is at most $\sim 207\ \mathrm{km}\ \mathrm{s}^{-1}$, or $|\Delta z| \leq 0.01$ \citep{bautista2017_bossBAOdatareduction} (approximately 1 COMAP channel). We note that this uncertainty does not account for the systematic gas in- or outflows that may be present in quasars; we discuss this additional consideration in \S\ref{sssec:ebossredshifts}.

We remove duplicates and any object with an SDSS \verb|NEGATIVE_EMISSION| warning from the eBOSS catalog, and then cut to the COMAP fields. In total, 243 eBOSS objects lie in the COMAP fields after these cuts. The average redshift of these objects is $z=2.73$, with their redshift distribution as shown in Figure \ref{fig:qso_redshift_distribution}. Their spatial distribution is shown in Figure \ref{fig:qso_distribution}, against the COMAP Season 1 sensitivity.

\section{Stacking Methods}\label{sec:stackmethods}

Broadly, the stack is a three-dimensional coaddition of the COMAP maps on the spatial and spectral positions of the quasars in the eBOSS catalog: we extract a three-dimensional `cubelet’ centered around the spatial position and redshifted CO(1-0) frequency of each source, and then combine them together into a single stacked cubelet containing the average luminosity of each source. 

\begin{figure*}[ht]
    \centering
    \includegraphics[width=0.8\textwidth]{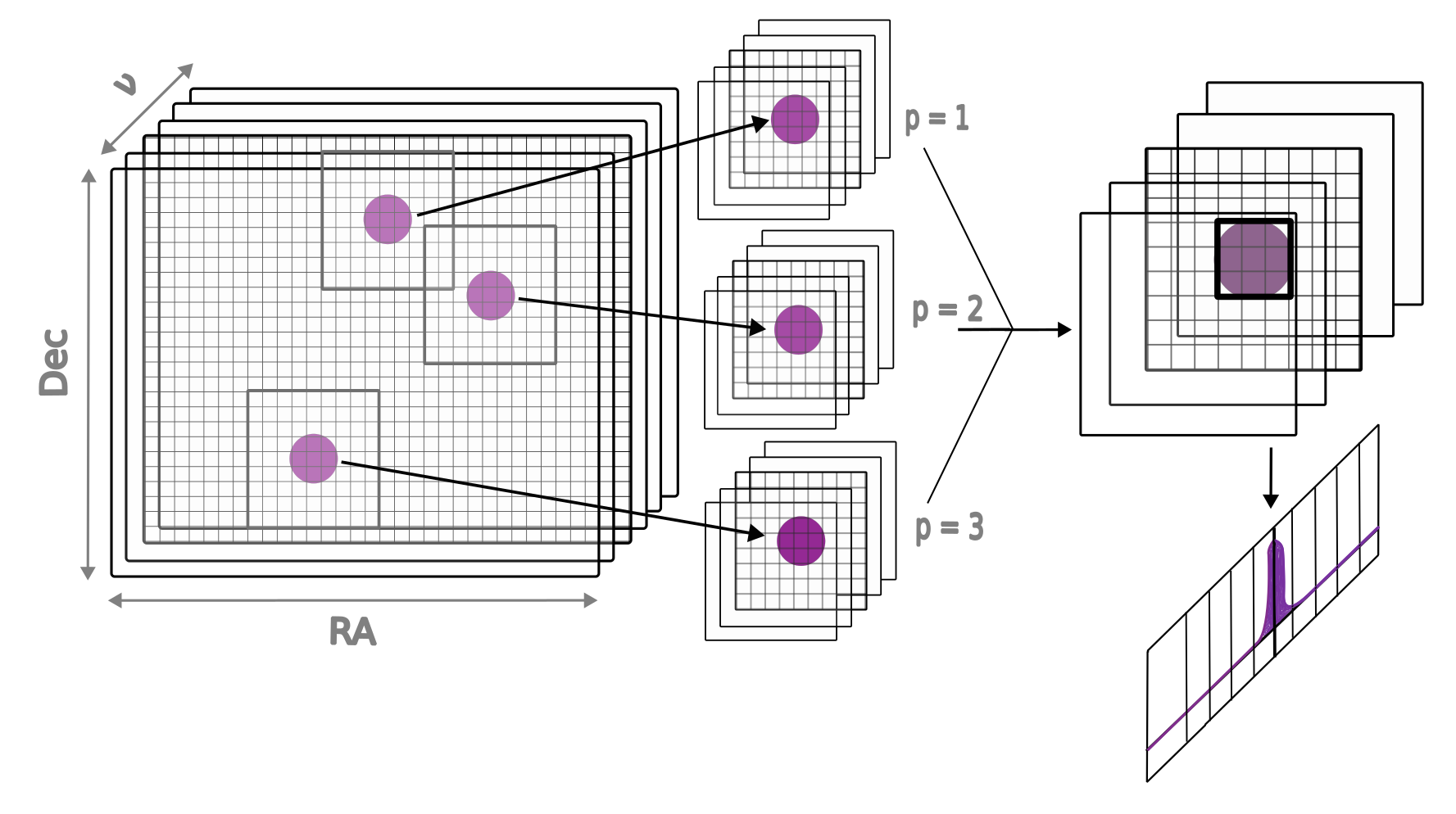}
    \caption{A diagram displaying the methodology used for this stacking analysis. From the full three-dimensional COMAP data cube (left), smaller 3-D cubelets are cut out centered on the position of each eBOSS object (center). These cubelets are then averaged into a single 3-D stack (top right), which is used to determine the average COMAP spectrum of the eBOSS objects (bottom right).}
    \label{fig:method_diagram}
\end{figure*}

\subsection{Determining Physical Quantities from the COMAP Data Cubes}\label{ssec:unit_conversions}
The COMAP data cubes are provided in brightness temperature units, $T_{\mathrm{b},i,j,k}$, where $i$ and $j$ index voxels in directions perpendicular to the line of sight (spatial axes) and $k$ indexes voxels parallel to the line of sight (spectral axis). The stacking pipeline begins by converting $T_{\mathrm{b},i,j,k}$ into velocity-integrated line flux, $(S\Delta v)_{i,j,k}$:
\begin{equation}
    (S\Delta v)_{i,j,k} = \frac{2\nu^2 k_B T_{\mathrm{b},i,j,k}}{c^2}\Omega_\mathrm{vox} \Delta v_{k}.
\end{equation} 
In this case, $T_{\mathrm{b},i,j,k}$ is the brightness temperature in the voxel, $\Omega_\mathrm{vox}$ is the solid angle of the sky subtended by the voxel (which is constant across the map), and $\Delta v_k$ is the channel width in km/s in the $k^\mathrm{th}$ frequency channel,
\begin{equation}
    \Delta v_k = \frac{\Delta \nu}{\nu_{\mathrm{obs},k}} c.
\end{equation}  
We follow \cite{Solomon_1997} to calculate \linelum\ in each voxel,
\begin{equation}
    L'_{\mathrm{CO},i,j,k}=\frac{c^2}{2k_B} (S \Delta v)_{i,j,k} \frac{D_{\mathrm{L},k}^2}{\nu_{\mathrm{obs},k}^{2}(1+z_k)^3},
\end{equation}
where $\nu_{\mathrm{obs},k}$ is the observed frequency of the $k^\mathrm{th}$ frequency channel, $z_k$ is the redshift required to place the 115.27 GHz CO(1-0) line into that channel's frequency range, and $D_{L,k}$ is the luminosity distance associated with that redshift. We perform this conversion before stacking because several of these values vary across the COMAP data cube.

\subsection{Extraction of Cubelets}\label{ssec:cubelet_cutouts}
For each eBOSS quasar, we extract from the converted COMAP \linelum\ cubes a three-dimensional `cubelet' centered around the spatial position and redshifted CO(1-0) frequency of the quasar. These cubelets have dimensions 42 arcmin $\times$ 42 arcmin $\times$ 1.28 GHz and contain CO line luminosity values $L'_{\mathrm{CO},{l,m,n,p}}$, where $l$ and $m$ are in spatial directions, $n$ is along the spectral axis, and $p$ indexes each cubelet. These are purposefully large compared to the COMAP spatial and spectral resolution, so they can be visually searched for potential large-scale fluctuations. If any cubelet has its central voxel masked, or more than half of the voxels immediately adjacent to the central voxel in all three directions masked, it is excluded from the catalog of cubelets to be stacked. 

\subsection{Stacking}\label{ssec:stacking}
We then stack the CO emission by directly combining the full three-dimensional cubelets associated with each individual galaxy, voxel-by-voxel. This is done to allow for additional testing of the fidelity of any potential detections. To account for the inhomogeneous noise response across the map (\S\ref{ssec:comap_data}), values are weighted by their RMS noise using inverse-variance weighting. Thus, the line luminosity in each voxel of the stacked cubelet is 
\begin{equation}
    L'_{\mathrm{CO},l,m,n} = \frac{\sum_p L'_{\mathrm{CO}, l,m,n,p} / \sigma^2_{l,m,n,p}}{\sum_{p} 1 / \sigma^2_{l,m,n,p}},
\end{equation}
\noindent for cubelet voxel line luminosities $L'_{\mathrm{CO},{l,m,n,p}}$ with RMS noise $\sigma_{l,m,n,p}$. The uncertainty in each voxel of the resulting stacked cubelet is then
\begin{equation}
    \sigma_{l,m,n} = \sqrt{\frac{1}{\sum_p 1 / \sigma^2_{l,m,n,p}}}.
\end{equation}
We additionally attempt to lessen any interactions between asymmetries in the position of catalog galaxies in the data cubes and the inhomogenous COMAP noise response by rotating each cubelet randomly in its spatial axes by either $0^\circ$, $90^\circ$, $180^\circ$, or $270^\circ$ before performing the coaddition. 

\subsection{Obtaining luminosity and density limits}\label{ssec:final_stack_values}
Once a full three-dimensional stack cubelet is obtained, we calculate the average line luminosity of the included quasars in two steps. Firstly, we calculate the line luminosity in each frequency channel, $L'_{\mathrm{CO},n}$, by summing over the central $3\time 3$ spatial pixels (spaxels) of the stacked cubelet (the region indicated with a black box in Figures \ref{fig:method_diagram} and \ref{fig:sim_stack_visual}),
\begin{equation}
    L'_{\mathrm{CO},n} = \sum_{l,m=-1}^{+1} L'_{\mathrm{CO}, l,m,n}
\end{equation}
which leaves us with a one-dimensional frequency spectrum. This $3\times3$ region corresponds to a $6'$ \rev{square} aperture, roughly 1.5 times the size of the COMAP main beam, meaning that most signal from any catalog object located anywhere in the central spaxel will not be spread outside this aperture by the beam. Thus, we avoid excluding any potential signal from the spectrum, while keeping the aperture as small as possible to mitigate the effects of beam dilution and maximize our signal-to-noise ratio (S/N). 

Finally, to determine the average CO luminosity of the stacked catalog objects $L'_\mathrm{CO}$, we sum the central seven frequency channels of the stacked spectrum,
\begin{equation}
    L'_\mathrm{CO} = \sum_{n=-3}^{+3} L'_{\mathrm{CO},n}.
\end{equation}
The seven-channel width corresponds to 218.75 MHz, or 2065 km/s at the average redshift of the galaxy catalog objects. This chosen spectral width allows for velocity offset between the CO emission and the optical lines used to determine the eBOSS redshifts, which can be quite significant (see \S\ref{sssec:ebossredshifts}). Instrumental aliasing between adjacent COMAP science channels is negligible \citep{lamb2021_instrument}, so broadening considerations in the spectral axes are almost entirely astrophysical.

We use the average line luminosity to determine an average molecular gas density in the stack. We use the \citet{bolatto13_cotoh2} CO-to-H$_2$ conversion factor of $3.6\ M_\odot$(K km s$^{-1}$ pc$^2$)$^{-1}$ (calculated for the Milky Way) to determine the molecular gas mass associated with our stacked line luminosity, and then convert this mass into a molecular gas density $\rho_\mathrm{H_2}$ by dividing by the \textit{comoving} volume of the stack aperture.

This is the simplest possible version of a stack, ignoring other available information such as the significance of the catalog objects' detections with eBOSS, or the optical luminosity of the catalog objects, both of which could also be used for weighting \citep[see, for example,][]{sinigaglia2022_symmstacking}. Functionally, this means assuming a $\delta$-function for the luminosity function. While we do plan to explore refinements to this methodology in the future, we assume for now that the noise in the Season 1 COMAP data cube is much greater than any variations across the eBOSS catalog.

\section{Data Verification}\label{sec:data_verification}
\subsection{Bootstrapped error analysis}
\label{ssec:bootstraps}
While we believe the noise in our maps to be Gaussian-distributed (`white' noise: see \S\ref{ssec:comap_data}, also \citeauthor{foss21_comapmapmaking}~\citeyear{foss21_comapmapmaking}) over the entire map, this stacking analysis introduces a new sampling strategy, and it is possible that the white noise assumption does not hold under this new sampling. We therefore perform a spatial `bootstrap' test on each stack to confirm the noise distribution in our maps, and to determine the uncertainty in each of the stacked values.

This is done by binning the actual eBOSS catalog by COMAP field and by redshift (into three redshift bins), and then generating an artificial catalog with the same overall redshift and field distribution as the real catalog but with random 3D positions. Variations in catalog number density between field, as well as instrumental effects such as the changing main beam with frequency and potential systematic errors correlating to on-sky position, will thus appear both in the bootstrapped stacks and the real stack. The LIM map is then stacked on the artificial catalog. This is repeated for 10000 different random artificial catalog iterations for each real stack run, to fully characterize the stack noise response.

\subsection{Constant-Luminosity Simulation Tests}\label{ssec:simulations}
\subsubsection{Simulation Methodology}\label{sssec:simulation_methodology}
As an end-to-end test of the stacking pipeline, we create mock COMAP observations of simulated cosmic CO signal, generate mock galaxy catalogs associated with these cubes, and perform the stacking analysis on these simulated data. Mock CO emission is generated by painting CO luminosities onto a dark matter halo catalog from peak-patch $N$-body simulations \citep{stein2019_peakpatchsims}. Because we are looking to check the efficacy of the stacking pipeline rather than make realistic cosmological predictions, we paint each halo with the same luminosity regardless of its mass. We artificially broaden the CO emission line in each halo by an effective velocity $v_\mathrm{eff}$ of 319 km/s (see \S\ref{sssec:ebossredshifts} below), again to more easily test the effects of the stacking pipeline. To generate mock COMAP data cubes, we additionally beam-smooth the emission in the spatial axes by convolving the mock cubes with a 2D Gaussian kernel with a 4-arcminute FWHM to approximate the COMAP main beam.

The artificial CO emission cube generated from these simulations is then converted into a timestream, scaled by a factor of 1000 to simulate an (unrealistically) strong detection, and added into real time-ordered data from the first COMAP observing season (Stutzer et al.,~\textit{in prep}). The positions of the simulated halos should have no correlation with any existing real cosmic structure, so this is an excellent way to simulate observations with the actual noise structure that COMAP observes. This mock timestream is passed through the COMAP pipeline, so any effects of the several pipeline filters \citep{foss21_comapmapmaking} will show up in the mock data cube.

In addition to the mock COMAP data cubes, we generate an artificial galaxy catalog by randomly selecting 1000 dark matter halos from the most massive 30\% of halos in the peak-patch catalog (dark-matter masses between $9\times10^{13}\ \mathrm{M}_\odot$ and $4\times10^{12}\ \mathrm{M}_\odot$), and taking these halos to be the objects emitting brightly enough in some other galaxy tracer to be detected in a traditional galaxy survey. Only the 3D positions of the halos are required for the stacking analysis, so we do not calculate any other mock parameters for the galaxy survey.

\begin{figure*}[ht!]
    \centering
    \includegraphics[width=\textwidth]{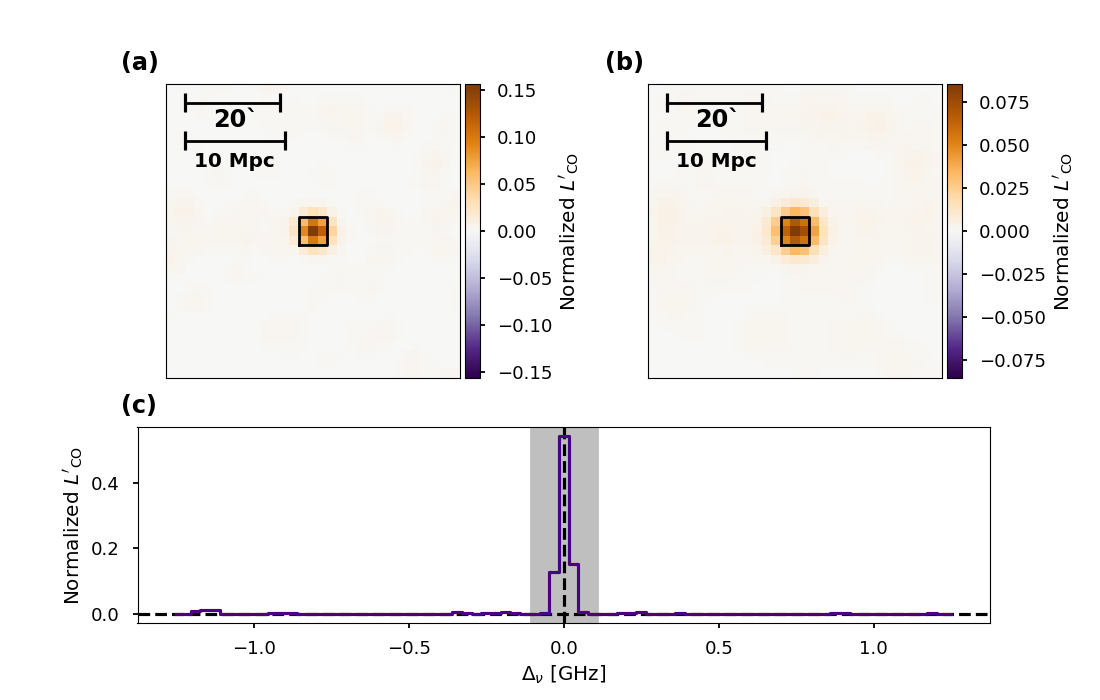}
    \caption{A \rev{noiseless} simulated stacking detection, normalized to the expected integrated luminosity. (a) The central three frequency channels coadded to provide a view of the spatial distribution of signal. The black box in the center indicates the smaller aperture used to calculate single stack values. (b) The same spatial representation of the stack, smoothed with a $4'$ Gaussian kernel to approximate the COMAP main beam. (c) The frequency spectrum of the stack. The value in each spectral channel is determined by summing over the spatial aperture indicated in (a). The three frequency channels making up the stack aperture are highlighted in grey. \rev{While the simulation is noiseless, neighbouring halos may show up in some objects' cutouts, appearing as small fluctuations in the stacked spectra.}}
    \label{fig:sim_stack_visual}
\end{figure*}

\subsubsection{Spatial attenuation due to COMAP beam}\label{sssec:simulation_results}

We then run the stacking pipeline on ten different realizations of the mock data cube and galaxy catalog. This is done principally to confirm that the COMAP pipeline is not attenuating the stacking signal in any unexpected ways, and to determine the size of the stacking aperture discussed in \S\ref{ssec:final_stack_values}. An example of a resulting simulated stack, shown in Figure \ref{fig:sim_stack_visual}, confirms that the $6'\times6'$ spatial aperture we selected indeed encloses the majority of the stacked flux, without incorporating much empty space. The line luminosity output by this stack is 88.5\% of the input value, meaning the input signal is attenuated by a factor of 1.13. This attenuation is due to the COMAP main beam spreading signal outside of the stack's $3\times3$ spatial aperture (Figure \ref{fig:sim_stack_visual}). Instead of broadening the aperture further and incorporating more empty space into the stack, which would affect our sensitivity, we multiply the values output by the stack by 1.13 to correct for this attenuation.

\subsubsection{Spectral attenuation due to offsets between optical and CO redshifts}\label{sssec:ebossredshifts}

Although the cataloged eBOSS redshifts are considered very precise, high-redshift quasars often contain large systematic outflows/inflows that cause velocity offsets between optical emission lines (such as those used by eBOSS), and the CO emission, which traces the cold gas of the galaxy itself \citep{banerji2017_qsoredshiftoffset, herreracamus2020_agnfeedbackinmerger, bischetti2021_apexredshifts_deltav_high-density}. This is an important consideration, as even small redshift uncertainties can cause significant attenuation in the signal of LIM/galaxy catalog cross-correlation analyses \citep{chung2019_crosscorrelation, chung21_linebroadening}. The shot component of the cross-power spectrum (\textit{i.e.},~the cross power at small scales, the most direct analogue to a stacking analysis) is where this effect is at its worst, although COMAP is most sensitive to larger spatial scales than those where attenuation due to redshift uncertainty is truly catastrophic.

\begin{figure}
    \centering
    \includegraphics[width=0.47\textwidth]{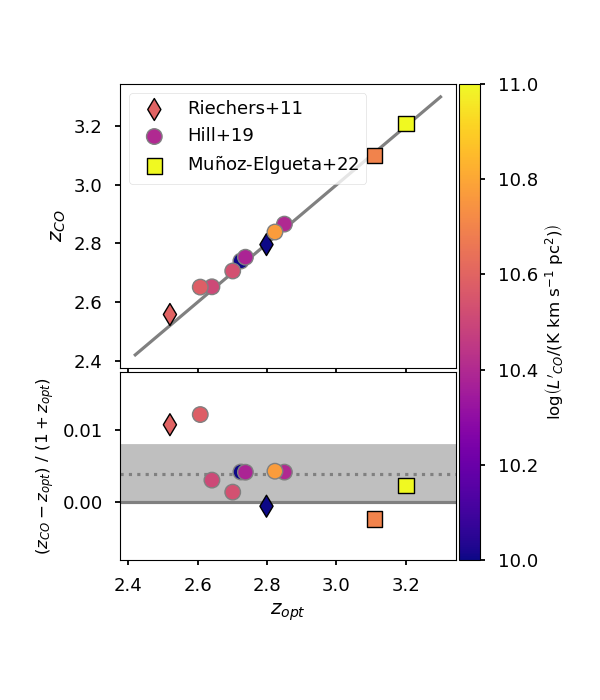}
    \caption{The optically-determined redshift values compared with CO-based redshifts for a selection of quasars individually detected in CO in the COMAP redshift range. The colourmap is the extrapolated CO(1-0) luminosity of each object, and the line where $z_\mathrm{opt}=z_\mathrm{CO}$ is shown for reference. In the bottom panel we indicate the mean offset of the sample with a dotted line, and the 1$\sigma$ scatter in that scatter with the shaded region.}
    \label{fig:eboss_redshift_comparison}
\end{figure}

To quanitfy systemic offsets between the cataloged redshift for our eBOSS sample (\zopt) and the redshift of the CO emission (\zco), we assemble a sample of quasars that have (clean) spectroscopic optical redshifts in the COMAP redshift range from SDSS, and have been individually detected in a molecular gas tracer line. These come from three studies:

\begin{itemize}
    \item Eight objects from \citeauthor{hill2019}~(\citeyear{hill2019}), who targeted CO(3-2) in 13 QSOs selected from the Keck Baryonic Structure Survey (KBSS) for additional cosmic web study with SCUBA-2. CO measurements used NOEMA \citep[the NOrthern Extended Millimeter Arary;][]{noema_pdbi}. The \zopt\ values for these objects are specifically from eBOSS. 
    
    \item Two objects from the sample of nine QSO MUSEUM quasars targeted for CO(6-5) and CO(7-6) observations with the SEPIA180 \citep{sepia180} receiver on APEX in \citeauthor{munozelgueta2022_apexqsos}~(\citeyear{munozelgueta2022_apexqsos}; ME22). These objects also have optical redshifts from eBOSS.
    
    \item Two quasars from \citeauthor{riechers2011_qsoCOlums}~(\citeyear{riechers2011_qsoCOlums}; R11), who targeted CO(1-0) in five quasar host galaxies using both the Extended Very Large Array (EVLA), and the Green Bank Telescope. Optical redshifts are from SDSS for these quasars.
    \end{itemize}
In each case, we exclude objects with no available eBOSS/SDSS redshift or that fall far outside the COMAP redshift range. We also exclude objects with SDSS \verb|NEGATIVE_EMISSION| warnings, for consistency with our own generated eBOSS catalog. Over this entire compiled catalog, the quasars have an average molecular gas emission-line FWHM of 319 km/s.

We plot \zopt\ against \zco\ for each of these quasars in Figure \ref{fig:eboss_redshift_comparison}. We find that \zopt\ is systematically smaller than \zco\, with a mean value of $(z_\mathrm{CO}-z_\mathrm{opt})/(1+z_\mathrm{opt}) = 0.00397 \pm 0.00408$. This offset does not appear to be correlated with either redshift or extrapolated CO(1-0) line luminosity. At the average redshift of our eBOSS quasar catalog, it corresponds to a negative frequency offset of $122$ MHz, or $1185$ km/s ($\sim 4$ COMAP spectral channels). This is in good agreement with observed bulk velocity offsets in quasars, which are typically $\sim 1000-2000$ km/s \citep{orellana2011_submmqso_redshifts, tytlerfan1992}.

In addition to exploring corrections in the bulk velocity offset between quasar gas phases, we also investigate the scatter in the relation. The standard deviation in $(z_\mathrm{CO}-z_\mathrm{opt}) / (1+z_\mathrm{opt})$ in the individually-detected quasar sample is $\sigma_{\delta z} = 0.00408$ from this effect alone, which is large enough to move a proportion of the signal from quasars in the stack outside of our defined 7-channel frequency aperture. We note that the variety of CO transitions used in the molecular gas measurements of the plotted quasars could create a larger scatter than a pure CO(1-0) analysis, as they trace slightly different phases of (dense) gas \citep[e.g.][]{munozelgueta2022_apexqsos}.

We perform a separate analysis to determine signal attenuation due to velocity offsets in quasars between the optical lines measured by eBOSS and the CO(1-0) line. This is addressed in detail in Appendix \ref{sec:spectral_attenuation_appendix}. We find that after this effect is accounted for, the input signal is attenuated by an additional factor of 1.58. As above, we multiply the values output by the stack by this factor to correct for this source of signal attenuation. Combined with the spatial attenuation discussed above, this is an overall attenuation factor of 1.79.

\section{Results}\label{sec:results}

\subsection{CO(1-0) Stack Results}

\begin{figure*}[ht!]
    \centering
    \includegraphics[width=\textwidth]{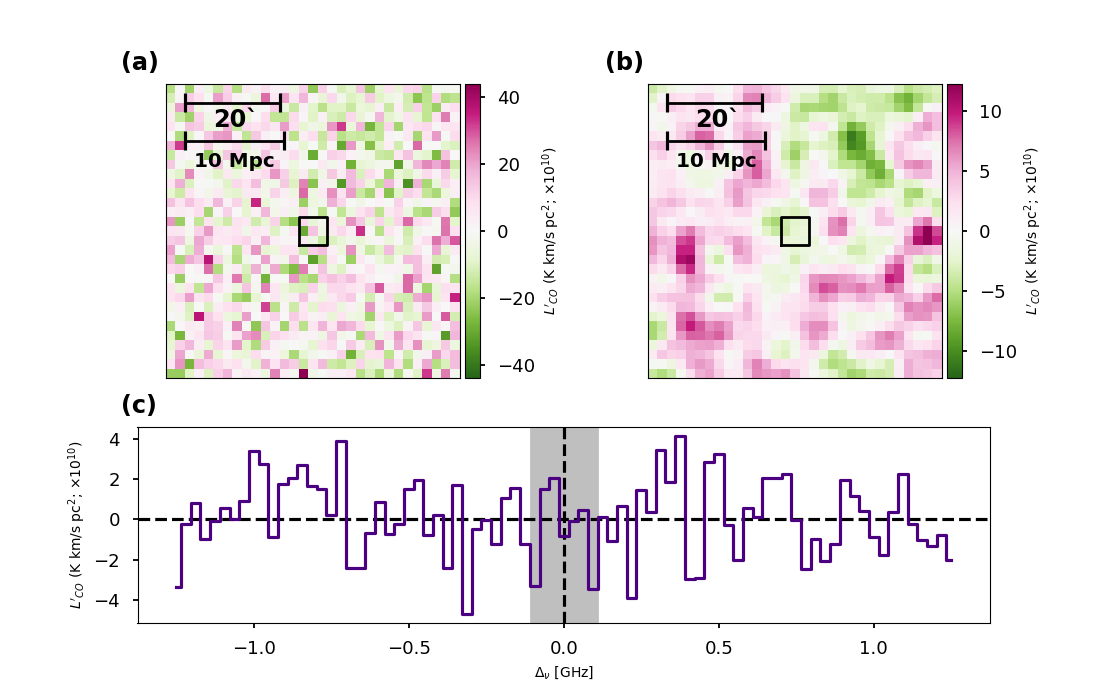}
    \caption{The result of stacking the COMAP data cubes on the 3D positions of eBOSS quasars. (a) The central seven frequency channels coadded to provide a view of the spatial distribution of signal. The black box in the center indicates the smaller aperture used to calculate single stack values. (b) The same spatial representation of the stack, smoothed with a $4'$ Gaussian kernel to approximate the COMAP main beam. (c) The frequency spectrum of the stack. The value in each spectral channel is determined by summing over the spatial aperture indicated in (a). The seven frequency channels making up the 3D stack aperture are highlighted in grey. All distance scales are given as proper values.}
    \label{fig:stack_visual}
\end{figure*}

\begin{figure}[h!]
    \centering
    \includegraphics[width=0.45\textwidth]{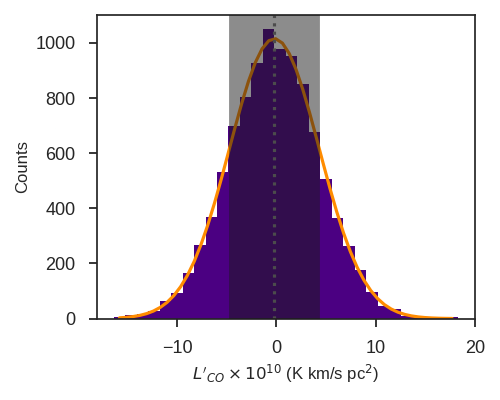}
    \caption{\rev{A histogram of the \linelum\ values returned from 10000 stacks performed on random 3D positions in the COMAP data cubes, with the same spectral and field distribution of objects as the real eBOSS catalog. A Gaussian fit to the histogram is shown in orange. The 66\% confidence interval of the fitted Gaussian (\textit{i.e.},~1 standard deviation about the fitted mean) is shown in dark gray, and the mean value is shown as a dotted gray line (see \S\ref{ssec:final_stack_values}).}}
    \label{fig:boss_bootstrap_test}
\end{figure}

The final COMAP/eBOSS stack is shown in Figure \ref{fig:stack_visual}. We find, at our current sensitivity level, no detection of CO emission associated with the 243 eBOSS quasars in the COMAP fields. \rev{The results of the COMAP/eBOSS bootstrap test (discussed in \S\ref{ssec:bootstraps}) are shown in Figure \ref{fig:boss_bootstrap_test}. As expected, the maps remain consistent with Gaussian noise centered at $T_\mathrm{b} = 0$ K even in this stacking analysis. As this is a more robust calculation of the distribution of stack values expected from random noise, we use the 95\% confidence region calculated from a Gaussian fit to these bootstrapped \linelum\ values as the uncertainty in our stack.}

We set a 95\% upper limit on the CO frequency-integrated line flux of $S\Delta v$ $\leq 0.129$ Jy km/s, corresponding to a limit on the CO line luminosity of \linelum $\ \leq 1.26\times 10^{11}\ $K km pc$^2$ s$^{-1}$ and a limit on the average molecular gas density in regions containing a quasar of \rhoh $\ \leq 1.52\times 10^{8}$ M$_{\odot}$ Mpc$^{-3}$. 

Additionally, as the quasar velocity offset discussed in Section \ref{sssec:ebossredshifts} is large enough to move any flux from our stack outside our chosen stack aperture, we correct for it by applying a 122 MHz frequency offset to our stack. We again find no significant CO emission, with a new upper limit of $L'_\mathrm{CO} \leq 8.69 \times10^{10}\ \mathrm{K\ km\ pc^2\ s^{-1}}$ (Figures \ref{fig:offset_stack_visual} and \ref{fig:offset_Bootstrap}). Accounting for this frequency offset thus does not meaningfully change our results, at our current level of sensitivity.  


\begin{table*}
\centering
\caption{95\% upper limits obtained by stacking the COMAP data cube on the CO(1-0) line and expected foreground emission lines. \label{tab:extralines}}
\begin{tabular}{r|c|c|c|c|c}
\hline
\hline
Spectral & \# of & $\nu_\mathrm{rest}$ & Redshift & $S\Delta v$ 95\% Upper Limit & $L'$ 95\% Upper Limit\\
Line & Objects & (GHz) & Range & (Jy km s$^{-1}$) & (K km pc$^2$ s$^{-1}$)\\
\hline
$^{12}$CO(1-0) & 243 & 115.27 & $2.4 - 3.4$ & $\leq 0.129$ & $\leq 1.26\times10^{11}$ \\
$^{12}$CO(1-0) with offset\tablenotemark{a} & 244 & 115.15 & $2.4 - 3.4$ & $\leq 0.270$ & $\leq 8.69\times10^{10}$ \\
\hline 
HCN(1-0) & 595 & 88.63 & $1.6 - 2.4$ & $\leq 0.196$ & $\leq 6.21\times10^{10}$ \\
CS(2-1) & 479 & 91.98 & $1.7 - 2.5$ & $\leq 0.285$ & $\leq 1.18\times10^{11}$  \\
$^{13}$CO(1-0) & 311 & 110.20 & $2.2 - 3.2$ & $\leq 0.236$ & $\leq 1.14\times10^{11}$ \\
CN$^-$(1-0) & 269 & 113.49 & $2.3 - 3.4$ & $\leq 0.316$ & $\leq 1.29\times10^{11}$ \\
\hline
\end{tabular}
\tablenotetext{a}{Shifted by the mean velocity offset of $z_\mathrm{eBOSS}$ from $z_\mathrm{CO}$ for the individually-detected quasars in \S\ref{sssec:ebossredshifts} (see Figure \ref{fig:eboss_redshift_comparison}).}
\end{table*}

\subsection{Foreground Lines}\label{ssec:foregrounds}
A major concern for LIM experiments is contamination of signal at the redshift of interest by other (primarily foreground) spectral lines, which may also be redshifted into the frequency range of the experiment by coincidence \citep{cheng2016_spectraldeconfusion, lidz2016_interlopercontamination}. CO LIM experiments suffer less from this type of contamination than LIM experiments targeting other spectral lines, because the (primarily molecular) lines that fall into our frequency range are expected to be faint compared to CO(1-0) \citep{cleary2021_comapoverview}. Joint analyses such as stacking, which require emission in two different spectral lines to be detected from the target sources, are also expected to be fairly robust against foreground contamination \citep{chung2019_crosscorrelation}. Because joint analyses are targeted toward specific emission lines, however, they present a unique opportunity for testing theoretical predictions that there will be minimal foreground emission present. Additionally, non-CO molecular emission lines are also interesting tracers of molecular gas in their own rights, and worth investigating.

Thus, we perform the stacking analysis on four other molecular emission lines which will likely fall into the COMAP frequency range, in addition to the $^{12}$CO(1-0) spectral line for which COMAP was designed. These lines are the $^{13}$CO isotope's $J=1\rightarrow 0$ rotational transition, the HCN $J=1\rightarrow 0$ transition, the CS $J=2\rightarrow 1$ transition, and the CN$^-$ radical's $J=1\rightarrow 0$ rotational transition. These lines are discussed in more detail in Appendix \ref{app:foregroundstacks}, or in \citet{breysse2015_foregrounds}. We make no detection of any foreground line. The resulting upper limits are shown in Table \ref{tab:extralines}, and representations of the stacked cubelets are shown in Appendix \ref{app:foregroundstacks}.

\subsection{Forecasts}\label{ssec:forecasts}

As we are working with only the first year of data from COMAP's planned multi-year Pathfinder survey, we also project the stack sensitivity after five years of observing. The eBOSS survey has finished, so we only forecast improvements to the stack sensitivity arising from deeper COMAP data, instead of other possible stack improvements such as including more catalog objects or refining their redshift or positional accuracy. Under this assumption, the stack will improve by an identical factor to the sensitivity of the COMAP data cubes themselves. 

Detailed projections were made by \cite{foss21_comapmapmaking} to forecast improvements to COMAP's sensitivity. Primarily, this improvement will be due to the obvious increase in total integration time by the end of the survey, but the Pathfinder's first season of operation was partially used for commissioning and refining the survey design, so we also expect significant improvements to the overall percentage of usable data. The effect of these improvements on COMAP's power spectrum sensitivity specifically are broken down in Table 2 of \citet{foss21_comapmapmaking}, and collated in Equation 26. Almost all of them apply to the stacking analysis, but some are power-spectrum specific:
\begin{itemize}
    \item The \cite{foss21_comapmapmaking} forecasts are made by projecting the sensitivity of a single COMAP field to the sensitivity of all three fields. As we are already using the full COMAP on-sky area in the stack analysis, we do not include the factors of $\sqrt{3}$ corresponding to this projection.
    
    \item Improvements to the filtering and map-making stages of the COMAP pipeline should result in these stages removing roughly 10\% less actual signal from the maps (this is the transfer function efficiency, $E_\mathrm{TF}$), but will mostly act on the large angular scales to which the stack is not sensitive. We therefore do not multiply by $E_\mathrm{TF}$ when forecasting for the stacks.
    
    \item The process of calculating the power spectrum from the completed intensity map involves generating many jackknifed cross-spectra and only keeping those that pass quality tests. The retention efficiency of these steps, parameterized by \citet{foss21_comapmapmaking} as $E_{\chi_{P(k)}^2}$ and $E_\mathrm{split}$, will likely improve in future COMAP seasons as the earlier pipeline stages become more robust. Since these data cuts use statistics calculated from already-generated intensity maps, they only affect the power spectrum itself and we do not include them here. 
\end{itemize}
All other efficiency improvements apply at the map level, and thus will apply to the stack. The forecast stacking sensitivity for the full COMAP Pathfinder survey then becomes
\begin{equation}
    \sigma^\mathrm{5yr}_\mathrm{stack} = \frac{\sigma^\mathrm{S1}_\mathrm{stack}}{\sqrt{D^\mathrm{5yr}_\mathrm{map}}},
\end{equation}
where
\begin{equation}
    D^\mathrm{5yr}_\mathrm{map} \equiv \frac{T^\mathrm{S1}E^\mathrm{S1}_\mathrm{tot} + (5\cdot 365 - T^\mathrm{S1}) E_\mathrm{tot}^\mathrm{proj} }{T^\mathrm{S1} E^\mathrm{S1}_\mathrm{tot} }.
\end{equation}
$T^\mathrm{S1}$ is the calendar duration of Season 1, 440 days. Both instances of $E_\mathrm{tot}$ indicate total observing efficiency, which combines bare on-sky observing efficiency $E_\mathrm{obs}$ with the fraction of data retained after all pipeline steps $E_\mathrm{data}$. For the exact breakdown of these quantities, see \cite{foss21_comapmapmaking}. $E_\mathrm{tot}^\mathrm{S1}$ is the actual Season 1 efficiency, and $E_\mathrm{tot}^\mathrm{proj}$ is the projected value after five years. Adjusting the values calculated by \cite{foss21_comapmapmaking} to exclude the power spectrum-only factors discussed above, these are 9.15\% and 33.2\%, respectively. We thus project that the sensitivity of the COMAP/eBOSS stack will improve by a factor of $D^\mathrm{5yr}_\mathrm{map} = 25.2$, corresponding to a CO(1-0) flux sensitivity $\sigma^\mathrm{5yr}_\mathrm{stack}$ of 0.026 Jy km s$^{-1}$ at the 95\% level.

While we have chosen in this work to focus specifically on the CO properties of quasars, we note that more general high-redshift spectroscopic catalogs can be used for stacking if the goal is to extract the CO properties of the universe as a whole. In particular, we have already made projections \citep{chung2019_crosscorrelation, chung2021_comapforecasts} for the cross-correlation of COMAP data with the HETDEX (the Hobby-Eberly Telescope Dark Energy EXperiment; \citeauthor{gebhardt2021_hetdexoverview}~\citeyear{gebhardt2021_hetdexoverview}) catalog of Lyman-Alpha Emitting galaxies (LAEs). We plan in future to perform a stacking analysis on the same catalog. To first order, the stack sensitivity increases with the number of objects $N$ in the spectroscopic catalog by a factor of $\sqrt{N}$, so we project significant sensitivity increases with this larger catalog. We note, however, that different galaxy tracers are subject to different biases, and any anticorrelation between the brightness of the tracers used and the CO brightness of a given object will affect this prediction; we plan to quantify this effect in future works.

\section{Discussion}\label{sec:discussion}

In general, how we interpret the results of LIM/galaxy survey stacking analyses depends on the relative contribution of the CO emission associated with the catalog galaxy compared to that of other sources within the COMAP beam. We can bracket the range of possible interpretations by considering the two extreme cases: i) the galaxies traced by our chosen survey dominate the brightness temperature of their surroundings (making the stack simply a measurement of the average CO luminosity, \linelum,\ of the survey objects), and ii) the cataloged galaxies are unspectacular themselves, but instead tend to fall into overdense regions of the large-scale structure (making the stack an measurement of the average CO luminosity of bright regions of the universe, which can be converted into their molecular gas density, \rhoh).

Due to the dearth of CO observations of quasars at cosmic noon, it is not yet clear which of these cases is closer to the truth; we discuss the available evidence below. Neither assumption, however, invalidates the other in the context of the upper limits we present here. Quasar companions will not be distinguishable from the objects themselves at COMAP resolutions, meaning any nearby objects will cause a source confusion effect pushing the measured \linelum\ value upwards. Conversely, abnormal amounts of CO emission from the quasars themselves will drive the local \rhoh\ upwards. In both cases, therefore, the stack values should likely be treated as upper limits even once a confident detection is made at this level of analysis.

\begin{figure*}[ht]
    \centering
    \includegraphics[width=0.95\textwidth]{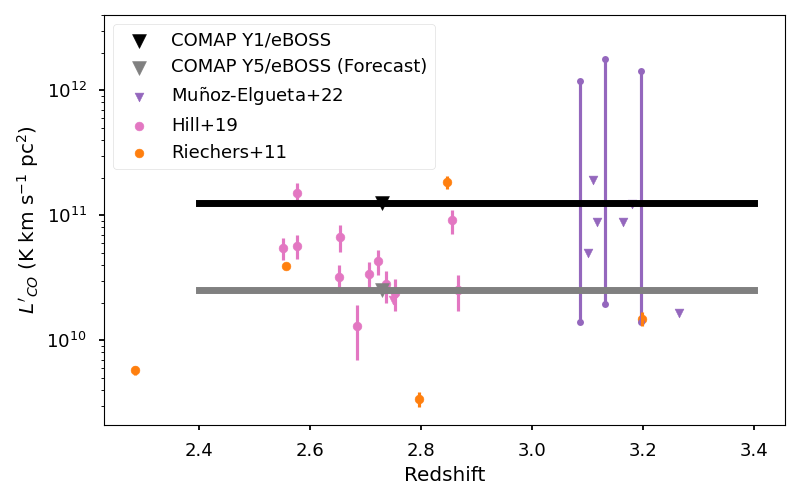}
    \caption{The COMAP/eBOSS 95\% upper limit on \linelum\ (the black bar), plotted against quasars detected individually in CO from \citet{riechers2011_qsoCOlums}, \citet{hill2019}, and \citet{munozelgueta2022_apexqsos}. \rev{Because \citet{munozelgueta2022_apexqsos} report a range of molecular gas mas values $M_\mathrm{H_2}$ from SED fitting of multiple lines, we show the \linelum\ values determined from those values as a range using vertical bars. Upper limits are shown as inverted triangles.} The extent of the COMAP/eBOSS bar indicates the redshift range to which COMAP is sensitive, and the marker is located at the mean redshift of the eBOSS objects on which we stack. A forecast for the stack sensitivity at the end of the full COMAP Pathfinder Survey is plotted in gray (see \S\ref{ssec:forecasts}).}
    \label{fig:line_lum_qsos}
\end{figure*}

\subsection{CO in Quasars}\label{ssec:quasarCOmeasurements}

The response of the molecular gas in quasar host galaxies to the extremely active SMBH at their centers is likely extremely complicated, especially at cosmic noon. The abundances of both quasars and gas-rich, sub-mm-bright galaxies (SMGs) seem to peak during this epoch, suggesting a link between the two \citep{simpson2012_smgs_and_qsos_connected}. Indeed, both theoretical and observational arguments have suggested that the SMBH activity driving quasars is triggered from normal SMGs by mergers rich in cold and cool gas \citep{sanders1988_galaxyqsolink, hopkins2008_mergersandquasars, brusa2018_COrichmergerQSO, herreracamus2020_agnfeedbackinmerger, bischetti2021_apexredshifts_deltav_high-density}, suggesting that, at least in their initial phases, quasars should be extremely rich in cold gas. 

Whether that cold gas survives the subsequent barrage of heat and pressure from its central SMBH, however, is still uncertain. SMG studies of molecular gas have shown that the presence of an AGN does not seem to significantly affect the gas properties of its host \citep{bothwell2013_moleculargas_smgs_agndoesntchange, banerji2017_qsoredshiftoffset}, and observed gas masses in quasars themselves follow typical SMG values closely \citep{hill2019}, but other observations have shown that CO Spectral-Line Energy Distributions (SLEDs) in quasar hosts tend towards higher excitation states than normal \citep[e.g.,][]{brusa2018_COrichmergerQSO}. At least for the purposes of COMAP, which measures the lowest CO excitation state, this would read as damped CO emission in quasar hosts. Additionally, many separate quasar studies have observed large-scale outflows of all gas phases from the host galaxy, although a fraction of this gas may remain in the molecular phase, and thus be observable with COMAP \citep[][]{bothwell2013_moleculargas_smgs_agndoesntchange, garciaburillo2014_moleculargasoutflow_A&A...567A.125G, feruglio2015_moleculargasoutflowA&A...583A..99F, brusa2018_COrichmergerQSO}. 

In the limiting case that much of the CO luminosity contributing to the LIM maps at the quasar positions comes from the quasars themselves, then the upper limit on \linelum\ determined from our COMAP/eBOSS stack becomes a direct upper limit on the average CO line luminosity of the 243 eBOSS quasars included in the stack. Under this analysis, the \linelum\ value can only ever be an upper limit, because the cosmic volume covered by the stack is large (9 Mpc $\times$ 9 Mpc $\times$ 2065 km/s\rev{, in proper units}) compared to an individual quasar, so it is unlikely that the quasars are the only sources contributing to the CO luminosity of the region. Because we are reporting an upper limit, however, this remains valid.

The measured average \linelum\ value is shown in Figure \ref{fig:line_lum_qsos}, plotted against other measures of quasar line luminosity in a similar redshift range, for comparison. These include the sample of 13 QSOs from KBSS, detected in CO(3-2) with NOEMA by \cite{hill2019}; the sample of 9 objects from the QSO MUSEUM, surveyed with APEX in CO(6-5) and CO(7-6) by \cite{munozelgueta2022_apexqsos}; and the sample of 5 QSOs detected in CO(1-0) by \citet{riechers2011_qsoCOlums} using the Extended Very Large Array and the Green Bank Telescope.

The \citet{hill2019} CO(3-2) measurements are converted into CO(1-0) line luminosities using the conversion factor $L'_\mathrm{CO(3-2)} / L'_\mathrm{CO(1-0)} = 0.97 \pm 0.19$ from \citet{carilliwalter2013_highzmoleculargas}. In several cases, a single KBSS object was found to be associated with two different CO sources slightly offset on-sky. For each of these objects, we sum both CO sources together to obtain the plotted \linelum\ value, as both sources would fall well into the COMAP beam. \citet{munozelgueta2022_apexqsos} converted their higher-$J$ CO measurements into molecular gas masses directly, by fitting to CO SLEDs, incorporating also the [CI] luminosity. In order to determine CO(1-0) line luminosities associated with the objects, we extrapolate their calculated $M_\mathrm{H_2}$ values to \linelum\ values using the Milky-Way $\alpha_\mathrm{CO}$ conversion factor of 3.6 M$_\odot$ ($\mathrm{K\ km\ s^{-1}\ pc^{-2}}$)$^{-1}$ \citep{bolatto13_cotoh2}, for consistency with our own previous analyses. These objects are mostly nondetections, and the SLED fitting returned a range of $M_\mathrm{H_2}$ values, so these ranges are what we plot in Figure \ref{fig:line_lum_qsos}. \citet{riechers2011_qsoCOlums} report CO(1-0) line luminosities directly. 

Figure \ref{fig:line_lum_qsos} shows that the COMAP/eBOSS stacking upper limit is already comparable with the line luminosities of (the brightest) individually surveyed quasars in its redshift range, even using only the first season of COMAP data. In some cases, the COMAP value is actually a stricter limit, particularly for the APEX objects \citep{munozelgueta2022_apexqsos}. This illustrates a powerful potential application for LIM -- directly detecting CO in individual objects becomes prohibitively expensive very quickly at these redshifts. A statistically-significant survey would require integration times that are not practical on high-demand community instruments. LIM-based stacking analyses, able to efficiently survey large samples of galaxies, could thus act as an important complement to these individual detection-based surveys by placing constraints on the ensemble emission from many galaxies. The minimum spatial scales accessible from a LIM instrument are large compared to galaxy sizes, meaning such analyses would have to remain upper limits even if a detection were made, but these still have the potential to be extremely enlightening.

Under this interpretation, our forecasted COMAP Pathfinder sensitivity to CO luminosity from a stack on the same 243 eBOSS quasars after five years of observing falls below the mean luminosity of the individually-detected quasars in Figure \ref{fig:line_lum_qsos}, meaning it has the potential to be genuinely constraining -- we will likely be able to tell with COMAP Y5 if these three samples are representative of the eBOSS quasar population as a whole. If we are able to constrain the \linelum\ of eBOSS quasars to this degree, it will provide significant insight into the complex feedback cycles present in quasars, and thus into star formation processes as a whole at cosmic noon.

\subsection{Cosmic Molecular Gas Density}\label{ssec:rhoh2}

\begin{figure*}
    \centering
    \includegraphics[width=\textwidth]{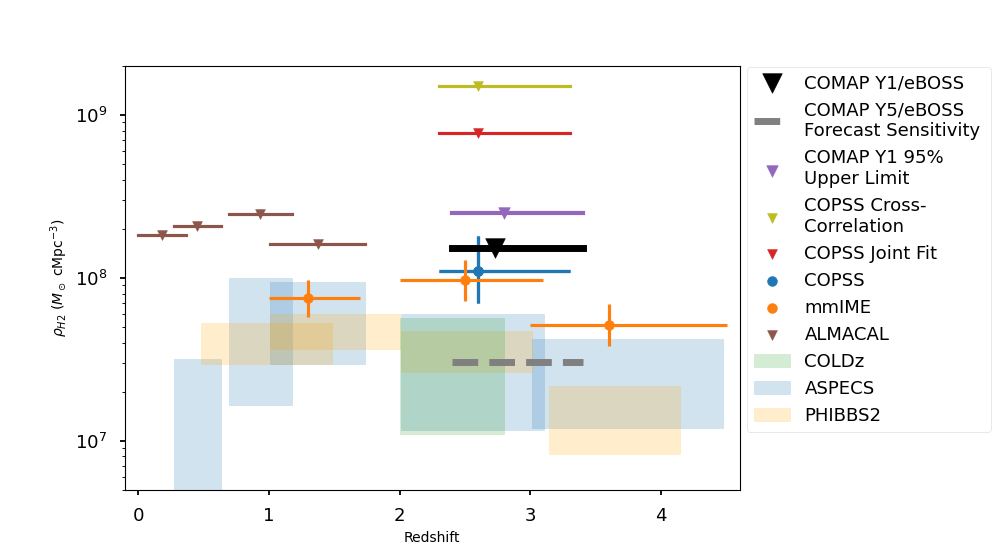}
    \caption{The upper limit on the average molecular gas density determined from the COMAP/eBOSS stack (black line), compared to other \rhoh\ measurements made using different survey techniques. The upper limit from the COMAP Season 1 power spectrum analysis is shown in purple. Also shown is the COPSS LIM survey's measured value \citep{keating2016_copss2}, as well as upper limits from COPSS joint analyses with galaxy surveys \citep{keenan2021_copssstack}. Shaded boxes are measurements made by large-scale traditional surveys COLDz \citep{riechers2019_coldz}, ASPECS \citep{gonzalezlopez2019_aspecs}, and PHIBSS2 \citep{lenkic2020_phibss2}. We show the forecast $2\sigma$ sensitivity level of the COMAP survey after five years stacked on eBOSS as a grey upper limit.}
    \label{fig:rhoh2_summary}
\end{figure*}

The other limiting case -- that quasars trace large-scale structure but do not themselves dominate the local CO emission -- is perhaps better-supported, as it follows from hierarchical structure formation. SMBH mass and galaxy mass have been shown several times to be correlated, meaning quasars will preferentially be found in massive galaxies that are likely to be at the center of large dark-matter halos \citep[e.g.,][]{gebhardt2000_smbh_galaxy_link, hopkins2008_mergersandquasars}. This argument is additionally supported by clustering measurements of both BOSS and SDSS Stripe 82 quasars \citep{white2012_quasarclustering_boss, timlin2018_quasarDMhalos_sdss}, as well as observations of the individual objects shown in Figure \ref{fig:line_lum_qsos}: a quarter of the \citet{hill2019} objects, for example, have nearby CO-bright companions \citep[also][]{banerji2017_qsoredshiftoffset, banerji2018_ismofreddenedquasars, decarli2021_moleculargas_qsoelans, bischetti2021_apexredshifts_deltav_high-density}. The argument that quasars require major mergers to ignite also necessitates at least one nearby massive galaxy in their recent past.

In this limit, where the quasar host itself is contributing negligibly to a local overdensity of CO emission, we can use the stack's upper limit on \tb\ to calculate the average molecular gas density \rhoh\ in bright regions at the COMAP redshifts. As the stack is only sensitive to a single, $\sim 9$ (comoving) Mpc spatial scale (the size of our chosen aperture), this value is most similar to a random `shot' power component of the \rhoh\ determined from a power spectrum analysis. We plot this value in Figure \ref{fig:rhoh2_summary}, alongside several other measurements of the same value made using a variety of techniques, including COMAP's own auto-spectrum based early science constraint \citep{chung2021_comapforecasts}. These other measurements include:

\begin{itemize}
    \item COPSS \citep{keating2016_copss2}, a LIM survey, targeting the same CO(1-0) transition as COMAP. COPSS used the Sunyaev-Zeldovich Array, and is subject to interferometric insensitivity to large scales. The \rhoh\ value from COPSS is thus based mainly on the shot-noise component of the CO power spectrum, converted to a molecular gas density using the Milky-Way $\alpha_\mathrm{CO}$ value of 3.6 M$_\odot$ (K km s$^{-1}$ pc$^{-2}$)$^{-1}$ (the same value we adopt in this work). Two other COPSS-determined values \citep{keenan2021_copssstack} are plotted, both determined using joint analyses combining COPSS data with (after correcting for main beam weighting) 145 spectroscopically-confirmed cataloged galaxies. \citet{keenan2021_copssstack} cross-correlated the COPSS data with a data cube obtained by gridding the catalog objects in 3D space (yielding one limit on \rhoh) and fit the resulting cross-power spectrum jointly with both the auto-power spectrum of COPSS alone and the galaxy-galaxy power spectrum of their galaxy catalog (yielding a second \rhoh\ limit).

    \item ALMACAL \citep{klitsch2019_almacal} used archival ALMA calibrators as background sources to search blindly for CO absorption lines. As these calibrator sources must be observed frequently, ALMACAL boasts $>1500$ hours of observations on a community instrument over a wide enough area to not be limited by cosmic variance. However, these calibrators do not extend past $z \sim 2$.
    
    \item The Millimeter-wave Intensity Mapping Experiment (mmIME) used a combination of archival and targeted ALMA and Atacama Compact Array (ACA) observations at frequencies tracing multiple CO transitions \citep{keating2020_mmime}. As with COPSS, these constraints are primarily constraints on the spectral shot power.

    \item Finally, the shaded boxes in Figure \ref{fig:rhoh2_summary} correspond to \rhoh\ measurements made using traditional (\textit{i.e.},~resolved and targeted) galaxy surveys. These include ASPECS \citep{uzgil2019_aspecs_pwrspec, gonzalezlopez2019_aspecs} and PHIBBS2 \citep{lenkic2020_phibss2}, each tracing multiple CO rotational transitions, as well as COLDz \citep{riechers2019_coldz}, which traces exclusively CO(1-0) at $z\sim 3$. While each of these traditional surveys provide \rhoh\ values that seem like much better constraints than any of the LIM-based numbers, they cover much smaller survey volumes and are thus extremely subject to cosmic varience effects. Additionally, as traditional surveys, they are insensitive to any emission below their detection limits, and so will miss any contribution to the total gas density from fainter objects, likely underestimating the total gas density.
\end{itemize}

Clearly, these \rhoh\ measurements were made using a diverse range of different techniques, so some of the scatter in Figure \ref{fig:rhoh2_summary} can be attributed simply to the biases associated with each strategy. There is also likely additional scatter present due to variations in the CO-to-H$_2$ ratio, both astrophysically and in the conversion factor $\alpha_\mathrm{CO}$ chosen for each analysis, as well as the relative intensity of the different CO line transitions, which will also change based on environment. 

In our case, while we are indeed subject to the assumptions associated with our choice of $\alpha_\mathrm{CO}$ and the bias of CO emission as a galaxy tracer, we are also only investigating the regions of space associated with eBOSS quasars, so the stacking analysis necessarily introduces a secondary bias factor. As discussed above, regions of this size surrounding quasars are likely to be more dense than the universe as a whole, biasing the stacked \rhoh\ value upwards because of clustering. The upper limit we report here is therefore likely an overestimate -- an analysis which fully explored clustering effects would be more constraining. Because we report an upper limit we find this to be acceptable. We will explore the effects of clustering more thoroughly in future work.

Due to this additional bias, the \rhoh\ value from the eBOSS stack is more similar to a cross-correlation analysis, such as the early WMAP cross-correlations with SDSS and BOSS quasars \citep{pullen2013_quasarsinWMAP, pullen2013_wmapquasars_limcrosscorr}. Unlike these cross-correlation analyses, the stack's \rhoh\ value is only sensitive to a single spatial scale (the scale of the 3D stack aperture, $\sim 9$ Mpc). We note that, because of our definition of a voxel in the COMAP data cubes, this scale does vary across the cube, and the \rhoh\ values being stacked are thus probing variable volumes. We explore this more thoroughly in Appendix \ref{sec:cosmic_volume_appendix}.

Even with these caveats, however, our predicted value is comparable to the values returned from the different COPSS joint analyses, reinforcing the viability of stacking as a LIM tool introduced by \cite{keenan2021_copssstack}. More promising is our forecast sensitivity for the COMAP Y5/eBOSS stack, which actually falls into the regions reported by traditional galaxy surveys. As these galaxy-survey values are likely underestimates of the actual cosmic H$_2$ density, our stacking analysis should detect emission from even very pessimistic models of cosmic CO, provided our forecasts hold true. As we already predict that our autocorrelation analysis will allow us to discriminate between CO models by COMAP Y5 \citep{chung2021_comapforecasts}, stacking will provide another valuable tool in characterizing the amount and properties of molecular gas emission at cosmic noon.

Additionally, quasar catalogs are far from the only spectroscopic information available on astrophysical objects at our redshifts. As discussed in \S\ref{ssec:forecasts}, increasing the number of objects in our galaxy catalog will increase the sensitivity of our stack, but introducing new objects detected using different tracers will also provide new opportunities for analysis. It will be very interesting to compare, for example, the CO properties of the LAEs in the HETDEX survey \citep{gebhardt2021_hetdexoverview} to the CO properties of the eBOSS quasars. Galaxy evolution is an inherently multitracer problem, so there will be significant analysis benefits to collating as many different surveys (whether traditional resolved surveys or LIM experiments) of the same region of space as possible. As the LIM field progresses, there will be more and more opportunities to do so.

\section{Summary and Conclusions}\label{sec:conclusion}

By stacking CO line-intensity maps from COMAP's first year of science operations on the 3D positions of quasars from the SDSS eBOSS survey, we have obtained a new upper limit on the cosmic abundance of molecular hydrogen gas. We describe the methodology behind the stack in detail, including the COMAP RMS-based weighting scheme we use when coadding and the bootstrapped error analysis technique we use to quantify uncertainties. We use the stack to search additionally for interloper emission from foreground galaxies in four molecular lines, finding no evidence of a detection in any case. We can interpret our stacked upper limit as a constraint on the CO emission from either the cataloged quasars themselves, or the regions of the universe surrounding those quasars. Likely, the most realistic interpretation is some combination of both cases. Additionally, we forecast stacking results for the COMAP Pathfinder survey after five years of observing.

In the limit that any potential signal in the stack is dominated by CO emission from the eBOSS quasars themselves, we treat our measurement as the average CO luminosity of these objects. We compare this average luminosity to resolved CO measurements of quasars in the COMAP redshift range, and find that our upper limit already probes CO luminosities fainter than those of the brightest objects observed to date. The quasar studies to which we compare typically required hours of observations using large community facilities to detect CO in only a handful of objects. Determining the CO properties of a large sample of objects such as our subset of the eBOSS catalog would thus be prohibitively expensive through traditional means. LIM stacks, therefore, could potentially be extremely useful as a tool for studying large samples of high-redshift galaxies and quasars.

Conversely, in the limit that the eBOSS quasars do not significantly contribute to the integrated CO emission of their surroundings, the stacked flux measurement can be converted to a measurement of the cosmic molecular gas density. While we compare this value directly to other $\rho_\mathrm{H_2}$ measurements from various sources, its interpretation is somewhat more complicated, as the molecular gas density in the regions traced by the stack depends heavily on the bias of the cataloged quasars towards large-scale structure. Additionally, the stack probes only a single spatial scale, unlike the more conventional power spectrum-based LIM measurements. These differences make stacking an excellent complement to other LIM analyses.

We propose, therefore, that stacking analyses with existing galaxy catalogs are a promising addition to the LIM analysis toolbox, especially when using LIM to approach as complex and multi-tracer a problem as galaxy evolution. To take full advantage of their potential benefits, stacking analyses should be performed on catalogs using as many different galaxy tracers as possible, to probe this phase space more fully. We aim to investigate stacking as a galaxy analysis tool more fully in future works, including by stacking on the extensive HETDEX catalog of Lyman-$\alpha$ emitters \citep{gebhardt2021_hetdexoverview}.

\section{Acknowledgements}

This material is based upon work supported by the National Science Foundation under Grant Nos. 1518282, 1517108, 1517598, 1517288, 1910999 and 2206834; by the Keck Institute for Space Studies under “The First Billion Years: A Technical Development Program for Spectral Line Observations”; and by a seed grant from the Kavli Institute for Particle Astrophysics and Cosmology.

DD acknowledges support from NSF Award 2206834. KC acknowledges support from NSF Awards 1518282, 1910999 and 2206834. PCB is supported by the James Arthur Postdoctoral Fellowship. DC is supported by a CITA/Dunlap Institute postdoctoral fellowship. The Dunlap Institute is funded through an endowment established by the David Dunlap family and the University of Toronto. HI, HKE, and IW acknowledge support from the Research Council of Norway through grant 251328. JG acknowledges support from the Keck Institute for Space Science, NSF AST-1517108 and University of Miami and Hugh Medrano for assistance with cryostat design. JK is supported by a Robert A. Millikan Fellowship from Caltech. HP's research is supported by the Swiss National Science Foundation via Ambizione Grant PZ00P2\_179934. 

This research made use of the SDSS-IV eBOSS survey. Funding for the Sloan Digital Sky 
Survey IV has been provided by the 
Alfred P. Sloan Foundation, the U.S. 
Department of Energy Office of 
Science, and the Participating 
Institutions. 

SDSS-IV acknowledges support and 
resources from the Center for High 
Performance Computing  at the 
University of Utah. The SDSS 
website is www.sdss4.org.

SDSS-IV is managed by the 
Astrophysical Research Consortium 
for the Participating Institutions 
of the SDSS Collaboration including 
the Brazilian Participation Group, 
the Carnegie Institution for Science, 
Carnegie Mellon University, Center for 
Astrophysics | Harvard \& 
Smithsonian, the Chilean Participation 
Group, the French Participation Group, 
Instituto de Astrof\'isica de 
Canarias, The Johns Hopkins 
University, Kavli Institute for the 
Physics and Mathematics of the 
Universe (IPMU) / University of 
Tokyo, the Korean Participation Group, 
Lawrence Berkeley National Laboratory, 
Leibniz Institut f\"ur Astrophysik 
Potsdam (AIP),  Max-Planck-Institut 
f\"ur Astronomie (MPIA Heidelberg), 
Max-Planck-Institut f\"ur 
Astrophysik (MPA Garching), 
Max-Planck-Institut f\"ur 
Extraterrestrische Physik (MPE), 
National Astronomical Observatories of 
China, New Mexico State University, 
New York University, University of 
Notre Dame, Observat\'ario 
Nacional / MCTI, The Ohio State 
University, Pennsylvania State 
University, Shanghai 
Astronomical Observatory, United 
Kingdom Participation Group, 
Universidad Nacional Aut\'onoma 
de M\'exico, University of Arizona, 
University of Colorado Boulder, 
University of Oxford, University of 
Portsmouth, University of Utah, 
University of Virginia, University 
of Washington, University of 
Wisconsin, Vanderbilt University, 
and Yale University.

This research made use of NASA’s Astrophysics Data System Bibliographic Services. For the purpose of open access, the authors have applied a Creative Commons Attribution (CC BY) licence to any Author Accepted Manuscript version arising from this submission. 

Finally, we would like to thank the anonymous referee whose comments and suggestions improved this manuscript.

\software{Astropy \citep{2013A&A...558A..33A,2018AJ....156..123A},
          Matplotlib \citep{Hunter:2007}}

\bibliography{bossstack}

\begin{thebibliography}{}
\expandafter\ifx\csname natexlab\endcsname\relax\def\natexlab#1{#1}\fi
\providecommand{\url}[1]{\href{#1}{#1}}
\providecommand{\dodoi}[1]{doi:~\href{http://doi.org/#1}{\nolinkurl{#1}}}
\providecommand{\doeprint}[1]{\href{http://ascl.net/#1}{\nolinkurl{http://ascl.net/#1}}}
\providecommand{\doarXiv}[1]{\href{https://arxiv.org/abs/#1}{\nolinkurl{https://arxiv.org/abs/#1}}}

\bibitem[{{Abdurashidova} {et~al.}(2022){Abdurashidova}, {Aguirre},
  {Alexander}, {Ali}, {Balfour}, {Beardsley}, {Bernardi}, {Billings}, {Bowman},
  {Bradley}, {Bull}, {Burba}, {Carey}, {Carilli}, {Cheng}, {DeBoer}, {Dexter},
  {de Lera Acedo}, {Dibblee-Barkman}, {Dillon}, {Ely}, {Ewall-Wice}, {Fagnoni},
  {Fritz}, {Furlanetto}, {Gale-Sides}, {Glendenning}, {Gorthi}, {Greig},
  {Grobbelaar}, {Halday}, {Hazelton}, {Hewitt}, {Hickish}, {Jacobs}, {Julius},
  {Kern}, {Kerrigan}, {Kittiwisit}, {Kohn}, {Kolopanis}, {Lanman}, {La Plante},
  {Lekalake}, {Lewis}, {Liu}, {MacMahon}, {Malan}, {Malgas}, {Maree},
  {Martinot}, {Matsetela}, {Mesinger}, {Molewa}, {Morales}, {Mosiane},
  {Murray}, {Neben}, {Nikolic}, {Nunhokee}, {Parsons}, {Patra}, {Pascua},
  {Pieterse}, {Pober}, {Razavi-Ghods}, {Ringuette}, {Robnett}, {Rosie}, {Sims},
  {Singh}, {Smith}, {Syce}, {Thyagarajan}, {Williams}, {Zheng}, \& {HERA
  Collaboration}}]{hera2021_earlyscience}
{Abdurashidova}, Z., {Aguirre}, J.~E., {Alexander}, P., {et~al.} 2022, \apj,
  925, 221, \dodoi{10.3847/1538-4357/ac1c78}

\bibitem[{Ahumada {et~al.}(2020)Ahumada, Prieto, Almeida, Anders, Anderson,
  Andrews, Anguiano, Arcodia, Armengaud, Aubert, {et~al.}}]{eboss_dr16}
Ahumada, R., Prieto, C.~A., Almeida, A., {et~al.} 2020, \apjs, 249, 3

\bibitem[{{Astropy Collaboration} {et~al.}(2013){Astropy Collaboration},
  {Robitaille}, {Tollerud}, {Greenfield}, {Droettboom}, {Bray}, {Aldcroft},
  {Davis}, {Ginsburg}, {Price-Whelan}, {Kerzendorf}, {Conley}, {Crighton},
  {Barbary}, {Muna}, {Ferguson}, {Grollier}, {Parikh}, {Nair}, {Unther},
  {Deil}, {Woillez}, {Conseil}, {Kramer}, {Turner}, {Singer}, {Fox}, {Weaver},
  {Zabalza}, {Edwards}, {Azalee Bostroem}, {Burke}, {Casey}, {Crawford},
  {Dencheva}, {Ely}, {Jenness}, {Labrie}, {Lim}, {Pierfederici}, {Pontzen},
  {Ptak}, {Refsdal}, {Servillat}, \& {Streicher}}]{2013A&A...558A..33A}
{Astropy Collaboration}, {Robitaille}, T.~P., {Tollerud}, E.~J., {et~al.} 2013,
  \aap, 558, A33, \dodoi{10.1051/0004-6361/201322068}

\bibitem[{{Astropy Collaboration} {et~al.}(2018){Astropy Collaboration},
  {Price-Whelan}, {Sip{\H{o}}cz}, {G{\"u}nther}, {Lim}, {Crawford}, {Conseil},
  {Shupe}, {Craig}, {Dencheva}, {Ginsburg}, {VanderPlas}, {Bradley},
  {P{\'e}rez-Su{\'a}rez}, {de Val-Borro}, {Aldcroft}, {Cruz}, {Robitaille},
  {Tollerud}, {Ardelean}, {Babej}, {Bach}, {Bachetti}, {Bakanov}, {Bamford},
  {Barentsen}, {Barmby}, {Baumbach}, {Berry}, {Biscani}, {Boquien}, {Bostroem},
  {Bouma}, {Brammer}, {Bray}, {Breytenbach}, {Buddelmeijer}, {Burke},
  {Calderone}, {Cano Rodr{\'\i}guez}, {Cara}, {Cardoso}, {Cheedella}, {Copin},
  {Corrales}, {Crichton}, {D'Avella}, {Deil}, {Depagne}, {Dietrich}, {Donath},
  {Droettboom}, {Earl}, {Erben}, {Fabbro}, {Ferreira}, {Finethy}, {Fox},
  {Garrison}, {Gibbons}, {Goldstein}, {Gommers}, {Greco}, {Greenfield},
  {Groener}, {Grollier}, {Hagen}, {Hirst}, {Homeier}, {Horton}, {Hosseinzadeh},
  {Hu}, {Hunkeler}, {Ivezi{\'c}}, {Jain}, {Jenness}, {Kanarek}, {Kendrew},
  {Kern}, {Kerzendorf}, {Khvalko}, {King}, {Kirkby}, {Kulkarni}, {Kumar},
  {Lee}, {Lenz}, {Littlefair}, {Ma}, {Macleod}, {Mastropietro}, {McCully},
  {Montagnac}, {Morris}, {Mueller}, {Mumford}, {Muna}, {Murphy}, {Nelson},
  {Nguyen}, {Ninan}, {N{\"o}the}, {Ogaz}, {Oh}, {Parejko}, {Parley}, {Pascual},
  {Patil}, {Patil}, {Plunkett}, {Prochaska}, {Rastogi}, {Reddy Janga},
  {Sabater}, {Sakurikar}, {Seifert}, {Sherbert}, {Sherwood-Taylor}, {Shih},
  {Sick}, {Silbiger}, {Singanamalla}, {Singer}, {Sladen}, {Sooley},
  {Sornarajah}, {Streicher}, {Teuben}, {Thomas}, {Tremblay}, {Turner},
  {Terr{\'o}n}, {van Kerkwijk}, {de la Vega}, {Watkins}, {Weaver}, {Whitmore},
  {Woillez}, {Zabalza}, \& {Astropy Contributors}}]{2018AJ....156..123A}
{Astropy Collaboration}, {Price-Whelan}, A.~M., {Sip{\H{o}}cz}, B.~M., {et~al.}
  2018, \aj, 156, 123, \dodoi{10.3847/1538-3881/aabc4f}

\bibitem[{{Banerji} {et~al.}(2017){Banerji}, {Carilli}, {Jones}, {Wagg},
  {McMahon}, {Hewett}, {Alaghband-Zadeh}, \&
  {Feruglio}}]{banerji2017_qsoredshiftoffset}
{Banerji}, M., {Carilli}, C.~L., {Jones}, G., {et~al.} 2017, \mnras, 465, 4390,
  \dodoi{10.1093/mnras/stw3019}

\bibitem[{{Banerji} {et~al.}(2018){Banerji}, {Jones}, {Wagg}, {Carilli},
  {Bisbas}, \& {Hewett}}]{banerji2018_ismofreddenedquasars}
{Banerji}, M., {Jones}, G.~C., {Wagg}, J., {et~al.} 2018, \mnras, 479, 1154,
  \dodoi{10.1093/mnras/sty1443}

\bibitem[{{Bautista} {et~al.}(2017){Bautista}, {Busca}, {Guy}, {Rich},
  {Blomqvist}, {du Mas des Bourboux}, {Pieri}, {Font-Ribera}, {Bailey},
  {Delubac}, {Kirkby}, {Le Goff}, {Margala}, {Slosar}, {Vazquez}, {Brownstein},
  {Dawson}, {Eisenstein}, {Miralda-Escud{\'e}}, {Noterdaeme},
  {Palanque-Delabrouille}, {P{\^a}ris}, {Petitjean}, {Ross}, {Schneider},
  {Weinberg}, \& {Y{\`e}che}}]{bautista2017_bossBAOdatareduction}
{Bautista}, J.~E., {Busca}, N.~G., {Guy}, J., {et~al.} 2017, \aap, 603, A12,
  \dodoi{10.1051/0004-6361/201730533}

\bibitem[{{Belitsky} {et~al.}(2018){Belitsky}, {Lapkin}, {Fredrixon},
  {Meledin}, {Sundin}, {Billade}, {Ferm}, {Pavolotsky}, {Rashid}, {Strandberg},
  {Desmaris}, {Ermakov}, {Krause}, {Olberg}, {Aghdam}, {Shafiee}, {Bergman},
  {De Beck}, {Olofsson}, {Conway}, {De Breuck}, {Immer}, {Yagoubov},
  {Montenegro-Montes}, {Torstensson}, {P{\'e}rez-Beaupuits}, {Klein}, {Boland},
  {Baryshev}, {Hesper}, {Barkhof}, {Adema}, {Bekema}, \& {Koops}}]{sepia180}
{Belitsky}, V., {Lapkin}, I., {Fredrixon}, M., {et~al.} 2018, \aap, 612, A23,
  \dodoi{10.1051/0004-6361/20173145810.48550/arXiv.1712.07396}

\bibitem[{{Bischetti} {et~al.}(2021){Bischetti}, {Feruglio}, {Piconcelli},
  {Duras}, {P{\'e}rez-Torres}, {Herrero}, {Venturi}, {Carniani}, {Bruni},
  {Gavignaud}, {Testa}, {Bongiorno}, {Brusa}, {Circosta}, {Cresci},
  {D'Odorico}, {Maiolino}, {Marconi}, {Mingozzi}, {Pappalardo}, {Perna},
  {Traianou}, {Travascio}, {Vietri}, {Zappacosta}, \&
  {Fiore}}]{bischetti2021_apexredshifts_deltav_high-density}
{Bischetti}, M., {Feruglio}, C., {Piconcelli}, E., {et~al.} 2021, \aap, 645,
  A33, \dodoi{10.1051/0004-6361/202039057}

\bibitem[{Bolatto {et~al.}(2013)Bolatto, Wolfire, \& Leroy}]{bolatto13_cotoh2}
Bolatto, A.~D., Wolfire, M., \& Leroy, A.~K. 2013, \araa, 51, 207,
  \dodoi{10.1146/annurev-astro-082812-140944}

\bibitem[{{Bolton} {et~al.}(2012){Bolton}, {Schlegel}, {Aubourg}, {Bailey},
  {Bhardwaj}, {Brownstein}, {Burles}, {Chen}, {Dawson}, {Eisenstein}, {Gunn},
  {Knapp}, {Loomis}, {Lupton}, {Maraston}, {Muna}, {Myers}, {Olmstead},
  {Padmanabhan}, {P{\^a}ris}, {Percival}, {Petitjean}, {Rockosi}, {Ross},
  {Schneider}, {Shu}, {Strauss}, {Thomas}, {Tremonti}, {Wake}, {Weaver}, \&
  {Wood-Vasey}}]{bolton2012_bossspectralreduction}
{Bolton}, A.~S., {Schlegel}, D.~J., {Aubourg}, {\'E}., {et~al.} 2012, \aj, 144,
  144, \dodoi{10.1088/0004-6256/144/5/144}

\bibitem[{{Bothwell} {et~al.}(2013){Bothwell}, {Smail}, {Chapman}, {Genzel},
  {Ivison}, {Tacconi}, {Alaghband-Zadeh}, {Bertoldi}, {Blain}, {Casey}, {Cox},
  {Greve}, {Lutz}, {Neri}, {Omont}, \&
  {Swinbank}}]{bothwell2013_moleculargas_smgs_agndoesntchange}
{Bothwell}, M.~S., {Smail}, I., {Chapman}, S.~C., {et~al.} 2013, \mnras, 429,
  3047, \dodoi{10.1093/mnras/sts562}

\bibitem[{Breysse \& Alexandroff(2019)}]{breysse2019_agnfeedback}
Breysse, P.~C., \& Alexandroff, R.~M. 2019, \mnras, 490, 260–273,
  \dodoi{10.1093/mnras/stz2534}

\bibitem[{{Breysse} {et~al.}(2015){Breysse}, {Kovetz}, \&
  {Kamionkowski}}]{breysse2015_foregrounds}
{Breysse}, P.~C., {Kovetz}, E.~D., \& {Kamionkowski}, M. 2015, \mnras, 452,
  3408, \dodoi{10.1093/mnras/stv1476}

\bibitem[{{Brusa} {et~al.}(2018){Brusa}, {Cresci}, {Daddi}, {Paladino},
  {Perna}, {Bongiorno}, {Lusso}, {Sargent}, {Casasola}, {Feruglio},
  {Fraternali}, {Georgiev}, {Mainieri}, {Carniani}, {Comastri}, {Duras},
  {Fiore}, {Mannucci}, {Marconi}, {Piconcelli}, {Zamorani}, {Gilli}, {La
  Franca}, {Lanzuisi}, {Lutz}, {Santini}, {Scoville}, {Vignali}, {Vito},
  {Rabien}, {Busoni}, \& {Bonaglia}}]{brusa2018_COrichmergerQSO}
{Brusa}, M., {Cresci}, G., {Daddi}, E., {et~al.} 2018, \aap, 612, A29,
  \dodoi{10.1051/0004-6361/201731641}

\bibitem[{Carilli \& Walter(2013)}]{carilliwalter2013_highzmoleculargas}
Carilli, C., \& Walter, F. 2013, Annual Review of Astronomy and Astrophysics,
  51, 105, \dodoi{10.1146/annurev-astro-082812-140953}

\bibitem[{{Cheng} {et~al.}(2016){Cheng}, {Chang}, {Bock}, {Bradford}, \&
  {Cooray}}]{cheng2016_spectraldeconfusion}
{Cheng}, Y.-T., {Chang}, T.-C., {Bock}, J., {Bradford}, C.~M., \& {Cooray}, A.
  2016, \apj, 832, 165, \dodoi{10.3847/0004-637X/832/2/165}

\bibitem[{Chenu {et~al.}(2016)Chenu, Navarrini, Bortolotti, Butin, Fontana,
  Mahieu, Maier, Mattiocco, Serres, Berton, Garnier, Moutote, Parioleau,
  Pissard, \& Reverdy}]{noema_pdbi}
Chenu, J.-Y., Navarrini, A., Bortolotti, Y., {et~al.} 2016, IEEE Transactions
  on Terahertz Science and Technology, 6, 223,
  \dodoi{10.1109/TTHZ.2016.2525762}

\bibitem[{{Chung} {et~al.}(2019){Chung}, {Viero}, {Church}, {Wechsler},
  {Alvarez}, {Bond}, {Breysse}, {Cleary}, {Eriksen}, {Foss}, {Gundersen},
  {Harper}, {Ihle}, {Keating}, {Murray}, {Padmanabhan}, {Stein}, {Wehus}, \&
  {COMAP Collaboration}}]{chung2019_crosscorrelation}
{Chung}, D.~T., {Viero}, M.~P., {Church}, S.~E., {et~al.} 2019, \apj, 872, 186,
  \dodoi{10.3847/1538-4357/ab0027}

\bibitem[{{Chung} {et~al.}(2021){Chung}, {Breysse}, {Ihle}, {Padmanabhan},
  {Silva}, {Bond}, {Borowska}, {Cleary}, {Eriksen}, {Foss}, {Gundersen},
  {Keating}, {Sturtzel Lunde}, {Philip}, {Stutzer}, {Viero}, {Watts}, {Wehus},
  \& {Wehus}}]{chung21_linebroadening}
{Chung}, D.~T., {Breysse}, P.~C., {Ihle}, H.~T., {et~al.} 2021, \apj, 923, 188,
  \dodoi{10.3847/1538-4357/ac2a35}

\bibitem[{{Chung} {et~al.}(2022){Chung}, {Breysse}, {Cleary}, {Ihle},
  {Padmanabhan}, {Silva}, {Richard Bond}, {Borowska}, {Catha}, {Church},
  {Dunne}, {Kristian Eriksen}, {Kristine Foss}, {Gaier}, {Ott Gundersen},
  {Harper}, {Harris}, {Hensley}, {Hobbs}, {Keating}, {Kim}, {Lamb}, {Lawrence},
  {Gahr Sturtzel Lunde}, {Murray}, {Pearson}, {Philip}, {Rasmussen},
  {Readhead}, {Rennie}, {Stutzer}, {Uzgil}, {Viero}, {Watts}, {Wechsler},
  {Kathrine Wehus}, {Woody}, \& {Comap
  Collaboration}}]{chung2021_comapforecasts}
{Chung}, D.~T., {Breysse}, P.~C., {Cleary}, K.~A., {et~al.} 2022, \apj, 933,
  186, \dodoi{10.3847/1538-4357/ac63c7}

\bibitem[{{Cleary} {et~al.}(2022){Cleary}, {Borowska}, {Breysse}, {Catha},
  {Chung}, {Church}, {Dickinson}, {Eriksen}, {Foss}, {Gundersen}, {Harper},
  {Harris}, {Hobbs}, {Ihle}, {Kim}, {Kocz}, {Lamb}, {Lunde}, {Padmanabhan},
  {Pearson}, {Philip}, {Powell}, {Rasmussen}, {Readhead}, {Rennie}, {Silva},
  {Stutzer}, {Uzgil}, {Watts}, {Wehus}, {Woody}, {Basoalto}, {Bond}, {Dunne},
  {Gaier}, {Hensley}, {Keating}, {Lawrence}, {Murray}, {Paladini}, {Reeves},
  {Viero}, {Wechsler}, \& {Comap Collaboration}}]{cleary2021_comapoverview}
{Cleary}, K.~A., {Borowska}, J., {Breysse}, P.~C., {et~al.} 2022, \apj, 933,
  182, \dodoi{10.3847/1538-4357/ac63cc}

\bibitem[{{Dawson} {et~al.}(2013){Dawson}, {Schlegel}, {Ahn}, {Anderson},
  {Aubourg}, {Bailey}, {Barkhouser}, {Bautista}, {Beifiori}, {Berlind},
  {Bhardwaj}, {Bizyaev}, {Blake}, {Blanton}, {Blomqvist}, {Bolton}, {Borde},
  {Bovy}, {Brandt}, {Brewington}, {Brinkmann}, {Brown}, {Brownstein}, {Bundy},
  {Busca}, {Carithers}, {Carnero}, {Carr}, {Chen}, {Comparat}, {Connolly},
  {Cope}, {Croft}, {Cuesta}, {da Costa}, {Davenport}, {Delubac}, {de Putter},
  {Dhital}, {Ealet}, {Ebelke}, {Eisenstein}, {Escoffier}, {Fan}, {Filiz Ak},
  {Finley}, {Font-Ribera}, {G{\'e}nova-Santos}, {Gunn}, {Guo}, {Haggard},
  {Hall}, {Hamilton}, {Harris}, {Harris}, {Ho}, {Hogg}, {Holder}, {Honscheid},
  {Huehnerhoff}, {Jordan}, {Jordan}, {Kauffmann}, {Kazin}, {Kirkby}, {Klaene},
  {Kneib}, {Le Goff}, {Lee}, {Long}, {Loomis}, {Lundgren}, {Lupton}, {Maia},
  {Makler}, {Malanushenko}, {Malanushenko}, {Mandelbaum}, {Manera}, {Maraston},
  {Margala}, {Masters}, {McBride}, {McDonald}, {McGreer}, {McMahon}, {Mena},
  {Miralda-Escud{\'e}}, {Montero-Dorta}, {Montesano}, {Muna}, {Myers},
  {Naugle}, {Nichol}, {Noterdaeme}, {Nuza}, {Olmstead}, {Oravetz}, {Oravetz},
  {Owen}, {Padmanabhan}, {Palanque-Delabrouille}, {Pan}, {Parejko},
  {P{\^a}ris}, {Percival}, {P{\'e}rez-Fournon}, {P{\'e}rez-R{\`a}fols},
  {Petitjean}, {Pfaffenberger}, {Pforr}, {Pieri}, {Prada}, {Price-Whelan},
  {Raddick}, {Rebolo}, {Rich}, {Richards}, {Rockosi}, {Roe}, {Ross}, {Ross},
  {Rossi}, {Rubi{\~n}o-Martin}, {Samushia}, {S{\'a}nchez}, {Sayres}, {Schmidt},
  {Schneider}, {Sc{\'o}ccola}, {Seo}, {Shelden}, {Sheldon}, {Shen}, {Shu},
  {Slosar}, {Smee}, {Snedden}, {Stauffer}, {Steele}, {Strauss}, {Streblyanska},
  {Suzuki}, {Swanson}, {Tal}, {Tanaka}, {Thomas}, {Tinker}, {Tojeiro},
  {Tremonti}, {Vargas Maga{\~n}a}, {Verde}, {Viel}, {Wake}, {Watson}, {Weaver},
  {Weinberg}, {Weiner}, {West}, {White}, {Wood-Vasey}, {Yeche}, {Zehavi},
  {Zhao}, \& {Zheng}}]{dawson2013_sdssbossreduction}
{Dawson}, K.~S., {Schlegel}, D.~J., {Ahn}, C.~P., {et~al.} 2013, \aj, 145, 10,
  \dodoi{10.1088/0004-6256/145/1/10}

\bibitem[{{Dawson} {et~al.}(2016){Dawson}, {Kneib}, {Percival}, {Alam},
  {Albareti}, {Anderson}, {Armengaud}, {Aubourg}, {Bailey}, {Bautista},
  {Berlind}, {Bershady}, {Beutler}, {Bizyaev}, {Blanton}, {Blomqvist},
  {Bolton}, {Bovy}, {Brandt}, {Brinkmann}, {Brownstein}, {Burtin}, {Busca},
  {Cai}, {Chuang}, {Clerc}, {Comparat}, {Cope}, {Croft}, \&
  {Cruz-Gonzalez}}]{2016AJ....151...44D}
{Dawson}, K.~S., {Kneib}, J.-P., {Percival}, W.~J., {et~al.} 2016, \apj, 151,
  44, \dodoi{10.3847/0004-6256/151/2/44}

\bibitem[{{Decarli} {et~al.}(2021){Decarli}, {Arrigoni-Battaia}, {Hennawi},
  {Walter}, {Prochaska}, \& {Cantalupo}}]{decarli2021_moleculargas_qsoelans}
{Decarli}, R., {Arrigoni-Battaia}, F., {Hennawi}, J.~F., {et~al.} 2021, \aap,
  645, L3, \dodoi{10.1051/0004-6361/202039814}

\bibitem[{{Feruglio} {et~al.}(2015){Feruglio}, {Fiore}, {Carniani},
  {Piconcelli}, {Zappacosta}, {Bongiorno}, {Cicone}, {Maiolino}, {Marconi},
  {Menci}, {Puccetti}, \&
  {Veilleux}}]{feruglio2015_moleculargasoutflowA&A...583A..99F}
{Feruglio}, C., {Fiore}, F., {Carniani}, S., {et~al.} 2015, \aap, 583, A99,
  \dodoi{10.1051/0004-6361/201526020}

\bibitem[{{Foss} {et~al.}(2022){Foss}, {Ihle}, {Borowska}, {Cleary}, {Eriksen},
  {Harper}, {Kim}, {Lamb}, {Lunde}, {Philip}, {Rasmussen}, {Stutzer}, {Uzgil},
  {Watts}, {Wehus}, {Woody}, {Bond}, {Breysse}, {Catha}, {Church}, {Chung},
  {Dickinson}, {Dunne}, {Gaier}, {Gundersen}, {Harris}, {Hobbs}, {Lawrence},
  {Murray}, {Readhead}, {Padmanabhan}, {Pearson}, {Rennie}, \& {Comap
  Collaboration}}]{foss21_comapmapmaking}
{Foss}, M.~K., {Ihle}, H.~T., {Borowska}, J., {et~al.} 2022, \apj, 933, 184,
  \dodoi{10.3847/1538-4357/ac63ca}

\bibitem[{{Gao} \& {Solomon}(2004)}]{gaosolomon2004_hcn}
{Gao}, Y., \& {Solomon}, P.~M. 2004, \apj, 606, 271, \dodoi{10.1086/382999}

\bibitem[{{Garc{\'\i}a-Burillo} {et~al.}(2014){Garc{\'\i}a-Burillo}, {Combes},
  {Usero}, {Aalto}, {Krips}, {Viti}, {Alonso-Herrero}, {Hunt}, {Schinnerer},
  {Baker}, {Boone}, {Casasola}, {Colina}, {Costagliola}, {Eckart}, {Fuente},
  {Henkel}, {Labiano}, {Mart{\'\i}n}, {M{\'a}rquez}, {Muller}, {Planesas},
  {Ramos Almeida}, {Spaans}, {Tacconi}, \& {van der
  Werf}}]{garciaburillo2014_moleculargasoutflow_A&A...567A.125G}
{Garc{\'\i}a-Burillo}, S., {Combes}, F., {Usero}, A., {et~al.} 2014, \aap, 567,
  A125, \dodoi{10.1051/0004-6361/201423843}

\bibitem[{{Gebhardt} {et~al.}(2000){Gebhardt}, {Bender}, {Bower}, {Dressler},
  {Faber}, {Filippenko}, {Green}, {Grillmair}, {Ho}, {Kormendy}, {Lauer},
  {Magorrian}, {Pinkney}, {Richstone}, \&
  {Tremaine}}]{gebhardt2000_smbh_galaxy_link}
{Gebhardt}, K., {Bender}, R., {Bower}, G., {et~al.} 2000, \apjl, 539, L13,
  \dodoi{10.1086/312840}

\bibitem[{{Gebhardt} {et~al.}(2021){Gebhardt}, {Mentuch Cooper}, {Ciardullo},
  {Acquaviva}, {Bender}, {Bowman}, {Castanheira}, {Dalton}, {Davis}, {de Jong},
  {DePoy}, {Devarakonda}, {Dongsheng}, {Drory}, {Fabricius}, {Farrow},
  {Feldmeier}, {Finkelstein}, {Froning}, {Gawiser}, {Gronwall}, {Herold},
  {Hill}, {Hopp}, {House}, {Janowiecki}, {Jarvis}, {Jeong}, {Jogee}, {Kakuma},
  {Kelz}, {Kollatschny}, {Komatsu}, {Krumpe}, {Landriau}, {Liu}, {Niemeyer},
  {MacQueen}, {Marshall}, {Mawatari}, {McLinden}, {Mukae}, {Nagaraj}, {Ono},
  {Ouchi}, {Papovich}, {Sakai}, {Saito}, {Schneider}, {Schulze},
  {Shanmugasundararaj}, {Shetrone}, {Sneden}, {Snigula}, {Steinmetz}, {Thomas},
  {Thomas}, {Tuttle}, {Urrutia}, {Wisotzki}, {Wold}, {Zeimann}, \&
  {Zhang}}]{gebhardt2021_hetdexoverview}
{Gebhardt}, K., {Mentuch Cooper}, E., {Ciardullo}, R., {et~al.} 2021, \apj,
  923, 217, \dodoi{10.3847/1538-4357/ac2e03}

\bibitem[{{Gonz{\'a}lez-L{\'o}pez} {et~al.}(2019){Gonz{\'a}lez-L{\'o}pez},
  {Decarli}, {Pavesi}, {Walter}, {Aravena}, {Carilli}, {Boogaard}, {Popping},
  {Weiss}, {Assef}, {Bauer}, {Bertoldi}, {Bouwens}, {Contini}, {Cortes}, {Cox},
  {da Cunha}, {Daddi}, {D{\'\i}az-Santos}, {Inami}, {Hodge}, {Ivison}, {Le
  F{\`e}vre}, {Magnelli}, {Oesch}, {Riechers}, {Rix}, {Smail}, {Swinbank},
  {Somerville}, {Uzgil}, \& {van der Werf}}]{gonzalezlopez2019_aspecs}
{Gonz{\'a}lez-L{\'o}pez}, J., {Decarli}, R., {Pavesi}, R., {et~al.} 2019, \apj,
  882, 139, \dodoi{10.3847/1538-4357/ab3105}

\bibitem[{{Gunn} {et~al.}(2006){Gunn}, {Siegmund}, {Mannery}, {Owen}, {Hull},
  {Leger}, {Carey}, {Knapp}, {York}, {Boroski}, {Kent}, {Lupton}, {Rockosi},
  {Evans}, {Waddell}, {Anderson}, {Annis}, {Barentine}, {Bartoszek}, {Bastian},
  {Bracker}, {Brewington}, {Briegel}, {Brinkmann}, {Brown}, {Carr},
  {Czarapata}, {Drennan}, {Dombeck}, {Federwitz}, {Gillespie}, {Gonzales},
  {Hansen}, {Harvanek}, {Hayes}, {Jordan}, {Kinney}, {Klaene}, {Kleinman},
  {Kron}, {Kresinski}, {Lee}, {Limmongkol}, {Lindenmeyer}, {Long}, {Loomis},
  {McGehee}, {Mantsch}, {Neilsen}, {Neswold}, {Newman}, {Nitta}, {Peoples},
  {Pier}, {Prieto}, {Prosapio}, {Rivetta}, {Schneider}, {Snedden}, \&
  {Wang}}]{gunn2006_sdsstelescope}
{Gunn}, J.~E., {Siegmund}, W.~A., {Mannery}, E.~J., {et~al.} 2006, \aj, 131,
  2332, \dodoi{10.1086/500975}

\bibitem[{{Herrera-Camus} {et~al.}(2020){Herrera-Camus}, {Janssen}, {Sturm},
  {Lutz}, {Veilleux}, {Davies}, {Shimizu}, {Gonz{\'a}lez-Alfonso}, {Rupke},
  {Tacconi}, {Genzel}, {Cicone}, {Maiolino}, {Contursi}, \&
  {Graci{\'a}-Carpio}}]{herreracamus2020_agnfeedbackinmerger}
{Herrera-Camus}, R., {Janssen}, A., {Sturm}, E., {et~al.} 2020, \aap, 635, A47,
  \dodoi{10.1051/0004-6361/201936434}

\bibitem[{{Hill} {et~al.}(2019){Hill}, {Chapman}, {Scott}, {Smail}, {Steidel},
  {Krips}, {Babul}, {Berg}, {Bertoldi}, {Gao}, {Lacaille}, {Matsuda}, {Ross},
  {Rudie}, \& {Trainor}}]{hill2019}
{Hill}, R., {Chapman}, S.~C., {Scott}, D., {et~al.} 2019, MNRAS, 485, 753,
  \dodoi{10.1093/mnras/stz429}

\bibitem[{Hinshaw {et~al.}(2013)Hinshaw, Larson, Komatsu, Spergel, Bennett,
  Dunkley, Nolta, Halpern, Hill, Odegard, Page, Smith, Weiland, Gold, Jarosik,
  Kogut, Limon, Meyer, Tucker, Wollack, \& Wright}]{Hinshaw_2013}
Hinshaw, G., Larson, D., Komatsu, E., {et~al.} 2013, \apjs, 208, 19,
  \dodoi{10.1088/0067-0049/208/2/19}

\bibitem[{{Hopkins} {et~al.}(2008){Hopkins}, {Hernquist}, {Cox}, \&
  {Kere{\v{s}}}}]{hopkins2008_mergersandquasars}
{Hopkins}, P.~F., {Hernquist}, L., {Cox}, T.~J., \& {Kere{\v{s}}}, D. 2008,
  \apjs, 175, 356, \dodoi{10.1086/524362}

\bibitem[{Hunter(2007)}]{Hunter:2007}
Hunter, J.~D. 2007, Computing in Science \& Engineering, 9, 90,
  \dodoi{10.1109/MCSE.2007.55}

\bibitem[{{Ihle} {et~al.}(2022){Ihle}, {Borowska}, {Cleary}, {Eriksen}, {Foss},
  {Harper}, {Kim}, {Lunde}, {Philip}, {Rasmussen}, {Stutzer}, {Uzgil}, {Watts},
  {Wehus}, {Bond}, {Breysse}, {Catha}, {Church}, {Chung}, {Dickinson}, {Dunne},
  {Gaier}, {Gundersen}, {Harris}, {Hobbs}, {Lamb}, {Lawrence}, {Murray},
  {Readhead}, {Padmanabhan}, {Pearson}, {Rennie}, {Woody}, \& {Compap
  Collaboration}}]{ihle2021_powerspectrum}
{Ihle}, H.~T., {Borowska}, J., {Cleary}, K.~A., {et~al.} 2022, \apj, 933, 185,
  \dodoi{10.3847/1538-4357/ac63c5}

\bibitem[{{Israel}(2020)}]{israel2020_13coratio_A&A...635A.131I}
{Israel}, F.~P. 2020, \aap, 635, A131, \dodoi{10.1051/0004-6361/201834198}

\bibitem[{{Jolly} {et~al.}(2021){Jolly}, {Knudsen}, {Laporte}, {Richard},
  {Fujimoto}, {Kohno}, {Ao}, {Bauer}, {Egami}, {Espada}, {Dessauges-Zavadsky},
  {Magdis}, {Schaerer}, {Sun}, {Valentino}, {Wang}, \&
  {Zitrin}}]{jolly2021_almaclusterstack}
{Jolly}, J.-B., {Knudsen}, K., {Laporte}, N., {et~al.} 2021, \aap, 652, A128,
  \dodoi{10.1051/0004-6361/202140878}

\bibitem[{{Keating} {et~al.}(2020){Keating}, {Marrone}, {Bower}, \&
  {Keenan}}]{keating2020_mmime}
{Keating}, G.~K., {Marrone}, D.~P., {Bower}, G.~C., \& {Keenan}, R.~P. 2020,
  \apj, 901, 141, \dodoi{10.3847/1538-4357/abb08e}

\bibitem[{Keating {et~al.}(2016)Keating, Marrone, Bower, Leitch, Carlstrom, \&
  DeBoer}]{keating2016_copss2}
Keating, G.~K., Marrone, D.~P., Bower, G.~C., {et~al.} 2016, \apj, 830, 34

\bibitem[{{Keenan} {et~al.}(2022){Keenan}, {Keating}, \&
  {Marrone}}]{keenan2021_copssstack}
{Keenan}, R.~P., {Keating}, G.~K., \& {Marrone}, D.~P. 2022, \apj, 927, 161,
  \dodoi{10.3847/1538-4357/ac4888}

\bibitem[{{Klitsch} {et~al.}(2019){Klitsch}, {P{\'e}roux}, {Zwaan}, {Smail},
  {Nelson}, {Popping}, {Chen}, {Diemer}, {Ivison}, {Allison}, {Muller},
  {Swinbank}, {Hamanowicz}, {Biggs}, \& {Dutta}}]{klitsch2019_almacal}
{Klitsch}, A., {P{\'e}roux}, C., {Zwaan}, M.~A., {et~al.} 2019, \mnras, 490,
  1220, \dodoi{10.1093/mnras/stz2660}

\bibitem[{{Kovetz} {et~al.}(2019){Kovetz}, {Breysse}, {Lidz}, {Bock},
  {Bradford}, {Chang}, {Foreman}, {Padmanabhan}, {Pullen}, {Riechers}, {Silva},
  \& {Switzer}}]{kovetz2019_limwhitepaper}
{Kovetz}, E., {Breysse}, P.~C., {Lidz}, A., {et~al.} 2019, \baas, 51, 101.
\newblock \doarXiv{1903.04496}

\bibitem[{{Lamb} {et~al.}(2022){Lamb}, {Cleary}, {Woody}, {Catha}, {Chung},
  {Gundersen}, {Harper}, {Harris}, {Hobbs}, {Ihle}, {Kocz}, {Pearson},
  {Philip}, {Powell}, {Basoalto}, {Bond}, {Borowska}, {Breysse}, {Church},
  {Dickinson}, {Dunne}, {Eriksen}, {Foss}, {Gaier}, {Kim}, {Lawrence}, {Lunde},
  {Padmanabhan}, {Rasmussen}, {Readhead}, {Reeves}, {Rennie}, {Stutzer},
  {Viero}, {Watts}, {Wehus}, \& {Comap Collaboration}}]{lamb2021_instrument}
{Lamb}, J.~W., {Cleary}, K.~A., {Woody}, D.~P., {et~al.} 2022, \apj, 933, 183,
  \dodoi{10.3847/1538-4357/ac63c6}

\bibitem[{{Lenki{\'c}} {et~al.}(2020){Lenki{\'c}}, {Bolatto}, {F{\"o}rster
  Schreiber}, {Tacconi}, {Neri}, {Combes}, {Walter}, {Garc{\'\i}a-Burillo},
  {Genzel}, {Lutz}, \& {Cooper}}]{lenkic2020_phibss2}
{Lenki{\'c}}, L., {Bolatto}, A.~D., {F{\"o}rster Schreiber}, N.~M., {et~al.}
  2020, \aj, 159, 190, \dodoi{10.3847/1538-3881/ab7458}

\bibitem[{{Lidz} \& {Taylor}(2016)}]{lidz2016_interlopercontamination}
{Lidz}, A., \& {Taylor}, J. 2016, \apj, 825, 143,
  \dodoi{10.3847/0004-637X/825/2/143}

\bibitem[{{Lujan Niemeyer} {et~al.}(2022){Lujan Niemeyer}, {Komatsu}, {Byrohl},
  {Davis}, {Fabricius}, {Gebhardt}, {Hill}, {Wisotzki}, {Bowman}, {Ciardullo},
  {Farrow}, {Finkelstein}, {Gawiser}, {Gronwall}, {Jeong}, {Landriau}, {Liu},
  {Cooper}, {Ouchi}, {Schneider}, \& {Zeimann}}]{lujanniemeyer2022}
{Lujan Niemeyer}, M., {Komatsu}, E., {Byrohl}, C., {et~al.} 2022, \apj, 929,
  90, \dodoi{10.3847/1538-4357/ac5cb8}

\bibitem[{Lyke {et~al.}(2020)Lyke, Higley, McLane, Schurhammer, Myers, Ross,
  Dawson, Chabanier, Martini, \& Busca}]{sdssdr16}
Lyke, B.~W., Higley, A.~N., McLane, J.~N., {et~al.} 2020, \apjs, 250, 8,
  \dodoi{10.3847/1538-4365/aba623}

\bibitem[{{Mu{\~n}oz-Elgueta} {et~al.}(2022){Mu{\~n}oz-Elgueta}, {Arrigoni
  Battaia}, {Kauffmann}, {De Breuck}, {Garc{\'\i}a-Vergara}, {Zanella},
  {Farina}, \& {Decarli}}]{munozelgueta2022_apexqsos}
{Mu{\~n}oz-Elgueta}, N., {Arrigoni Battaia}, F., {Kauffmann}, G., {et~al.}
  2022, \mnras, 511, 1462, \dodoi{10.1093/mnras/stac041}

\bibitem[{{Orellana} {et~al.}(2011){Orellana}, {Nagar}, {Isaak}, {Priddey},
  {Maiolino}, {McMahon}, {Marconi}, \&
  {Oliva}}]{orellana2011_submmqso_redshifts}
{Orellana}, G., {Nagar}, N.~M., {Isaak}, K.~G., {et~al.} 2011, \aap, 531, A128,
  \dodoi{10.1051/0004-6361/201015807}

\bibitem[{{Paul} {et~al.}(2023){Paul}, {Santos}, {Chen}, \&
  {Wolz}}]{paul2023_meerkat}
{Paul}, S., {Santos}, M.~G., {Chen}, Z., \& {Wolz}, L. 2023, arXiv e-prints,
  arXiv:2301.11943, \dodoi{10.48550/arXiv.2301.11943}

\bibitem[{{Pullen} {et~al.}(2013){Pullen}, {Chang}, {Dor{\'e}}, \&
  {Lidz}}]{pullen2013_wmapquasars_limcrosscorr}
{Pullen}, A.~R., {Chang}, T.-C., {Dor{\'e}}, O., \& {Lidz}, A. 2013, \apj, 768,
  15, \dodoi{10.1088/0004-637X/768/1/15}

\bibitem[{{Pullen} \& {Hirata}(2013)}]{pullen2013_quasarsinWMAP}
{Pullen}, A.~R., \& {Hirata}, C.~M. 2013, \pasp, 125, 705,
  \dodoi{10.1086/671189}

\bibitem[{{Rennie} {et~al.}(2022){Rennie}, {Harper}, {Dickinson}, {Philip},
  {Cleary}, {Bond}, {Borowska}, {Breysse}, {Catha}, {Cepeda-Arroita}, {Chung},
  {Church}, {Dunne}, {Eriksen}, {Foss}, {Gaier}, {Gundersen}, {Harris},
  {Hensley}, {Hobbs}, {Ihle}, {Lamb}, {Lawrence}, {Lunde}, {Paladini},
  {Pearson}, {Rasmussen}, {Readhead}, {Stutzer}, {Watts}, {Wehus}, {Woody}, \&
  {Comap Collaboration}}]{rennie2022_comapgps}
{Rennie}, T.~J., {Harper}, S.~E., {Dickinson}, C., {et~al.} 2022, \apj, 933,
  187, \dodoi{10.3847/1538-4357/ac63c8}

\bibitem[{{Riechers} {et~al.}(2007){Riechers}, {Walter}, {Cox}, {Carilli},
  {Weiss}, {Bertoldi}, \& {Neri}}]{riechers2007_cn_ApJ...666..778R}
{Riechers}, D.~A., {Walter}, F., {Cox}, P., {et~al.} 2007, \apj, 666, 778,
  \dodoi{10.1086/520335}

\bibitem[{{Riechers} {et~al.}(2011){Riechers}, {Carilli}, {Maddalena}, {Hodge},
  {Harris}, {Baker}, {Walter}, {Wagg}, {Vanden Bout}, {Wei{\ss}}, \&
  {Sharon}}]{riechers2011_qsoCOlums}
{Riechers}, D.~A., {Carilli}, C.~L., {Maddalena}, R.~J., {et~al.} 2011, \apjl,
  739, L32, \dodoi{10.1088/2041-8205/739/1/L32}

\bibitem[{{Riechers} {et~al.}(2019){Riechers}, {Pavesi}, {Sharon}, {Hodge},
  {Decarli}, {Walter}, {Carilli}, {Aravena}, {da Cunha}, {Daddi}, {Dickinson},
  {Smail}, {Capak}, {Ivison}, {Sargent}, {Scoville}, \&
  {Wagg}}]{riechers2019_coldz}
{Riechers}, D.~A., {Pavesi}, R., {Sharon}, C.~E., {et~al.} 2019, \apj, 872, 7,
  \dodoi{10.3847/1538-4357/aafc27}

\bibitem[{{Romano} {et~al.}(2022){Romano}, {Morselli}, {Cassata}, {Ginolfi},
  {Schaerer}, {B{\'e}thermin}, {Capak}, {Faisst}, {Le F{\`e}vre}, {Silverman},
  {Yan}, {Bardelli}, {Boquien}, {Dessauges-Zavadsky}, {Fujimoto}, {Hathi},
  {Jones}, {Koekemoer}, {Lemaux}, {M{\'e}ndez-Hern{\'a}ndez}, {Narayanan},
  {Talia}, {Vergani}, {Zamorani}, \& {Zucca}}]{romano2022_alpinestack}
{Romano}, M., {Morselli}, L., {Cassata}, P., {et~al.} 2022, \aap, 660, A14,
  \dodoi{10.1051/0004-6361/202142265}

\bibitem[{{Sanders} {et~al.}(1988){Sanders}, {Soifer}, {Elias}, {Neugebauer},
  \& {Matthews}}]{sanders1988_galaxyqsolink}
{Sanders}, D.~B., {Soifer}, B.~T., {Elias}, J.~H., {Neugebauer}, G., \&
  {Matthews}, K. 1988, \apjl, 328, L35, \dodoi{10.1086/185155}

\bibitem[{{Silva} {et~al.}(2021){Silva}, {Baumschlager}, {Cleary}, {Breysse},
  {Chung}, {Ihle}, {Padmanabhan}, {Keating}, {Kim}, \&
  {Philip}}]{silva2021_hetdexcomap}
{Silva}, M.~B., {Baumschlager}, B., {Cleary}, K.~A., {et~al.} 2021, arXiv
  e-prints, arXiv:2111.05354.
\newblock \doarXiv{2111.05354}

\bibitem[{{Simpson} {et~al.}(2012){Simpson}, {Smail}, {Swinbank}, {Alexander},
  {Auld}, {Baes}, {Bonfield}, {Clements}, {Cooray}, {Coppin}, {Danielson},
  {Dariush}, {Dunne}, {de Zotti}, {Harrison}, {Hopwood}, {Hoyos}, {Ibar},
  {Ivison}, {Jarvis}, {Lapi}, {Maddox}, {Page}, {Riechers}, {Valiante}, \& {van
  der Werf}}]{simpson2012_smgs_and_qsos_connected}
{Simpson}, J.~M., {Smail}, I., {Swinbank}, A.~M., {et~al.} 2012, \mnras, 426,
  3201, \dodoi{10.1111/j.1365-2966.2012.21941.x}

\bibitem[{{Sinigaglia} {et~al.}(2022){Sinigaglia}, {Elson}, {Rodighiero}, \&
  {Vaccari}}]{sinigaglia2022_symmstacking}
{Sinigaglia}, F., {Elson}, E., {Rodighiero}, G., \& {Vaccari}, M. 2022, \mnras,
  514, 4205, \dodoi{10.1093/mnras/stac1584}

\bibitem[{{Smee} {et~al.}(2013){Smee}, {Gunn}, {Uomoto}, {Roe}, {Schlegel},
  {Rockosi}, {Carr}, {Leger}, {Dawson}, {Olmstead}, {Brinkmann}, {Owen},
  {Barkhouser}, {Honscheid}, {Harding}, {Long}, {Lupton}, {Loomis}, {Anderson},
  {Annis}, {Bernardi}, {Bhardwaj}, {Bizyaev}, {Bolton}, {Brewington}, {Briggs},
  {Burles}, {Burns}, {Castander}, {Connolly}, {Davenport}, {Ebelke}, {Epps},
  {Feldman}, {Friedman}, {Frieman}, {Heckman}, {Hull}, {Knapp}, {Lawrence},
  {Loveday}, {Mannery}, {Malanushenko}, {Malanushenko}, {Merrelli}, {Muna},
  {Newman}, {Nichol}, {Oravetz}, {Pan}, {Pope}, {Ricketts}, {Shelden},
  {Sandford}, {Siegmund}, {Simmons}, {Smith}, {Snedden}, {Schneider},
  {SubbaRao}, {Tremonti}, {Waddell}, \& {York}}]{smee2013_bossspectrograph}
{Smee}, S.~A., {Gunn}, J.~E., {Uomoto}, A., {et~al.} 2013, \aj, 146, 32,
  \dodoi{10.1088/0004-6256/146/2/32}

\bibitem[{Solomon {et~al.}(1997)Solomon, Downes, Radford, \&
  Barrett}]{Solomon_1997}
Solomon, P.~M., Downes, D., Radford, S. J.~E., \& Barrett, J.~W. 1997, \apj,
  478, 144, \dodoi{10.1086/303765}

\bibitem[{{Stanley} {et~al.}(2019){Stanley}, {Jolly}, {K{\"o}nig}, \&
  {Knudsen}}]{stanley2019_quasarstack}
{Stanley}, F., {Jolly}, J.~B., {K{\"o}nig}, S., \& {Knudsen}, K.~K. 2019, \aap,
  631, A78, \dodoi{10.1051/0004-6361/201834530}

\bibitem[{{Stein} {et~al.}(2019){Stein}, {Alvarez}, \&
  {Bond}}]{stein2019_peakpatchsims}
{Stein}, G., {Alvarez}, M.~A., \& {Bond}, J.~R. 2019, \mnras, 483, 2236,
  \dodoi{10.1093/mnras/sty3226}

\bibitem[{{Timlin} {et~al.}(2018){Timlin}, {Ross}, {Richards}, {Myers},
  {Pellegrino}, {Bauer}, {Lacy}, {Schneider}, {Wollack}, \&
  {Zakamska}}]{timlin2018_quasarDMhalos_sdss}
{Timlin}, J.~D., {Ross}, N.~P., {Richards}, G.~T., {et~al.} 2018, \apj, 859,
  20, \dodoi{10.3847/1538-4357/aab9ac}

\bibitem[{{Tytler} \& {Fan}(1992)}]{tytlerfan1992}
{Tytler}, D., \& {Fan}, X.-M. 1992, \apjs, 79, 1, \dodoi{10.1086/191642}

\bibitem[{{Uzgil} {et~al.}(2019){Uzgil}, {Carilli}, {Lidz}, {Walter},
  {Thyagarajan}, {Decarli}, {Aravena}, {Bertoldi}, {Cortes},
  {Gonz{\'a}lez-L{\'o}pez}, {Inami}, {Popping}, {Riechers}, {Van der Werf},
  {Wagg}, \& {Weiss}}]{uzgil2019_aspecs_pwrspec}
{Uzgil}, B.~D., {Carilli}, C., {Lidz}, A., {et~al.} 2019, \apj, 887, 37,
  \dodoi{10.3847/1538-4357/ab517f}

\bibitem[{{Wang} {et~al.}(2011){Wang}, {Zhang}, \&
  {Shi}}]{wang2011_cs_MNRAS.416L..21W}
{Wang}, J., {Zhang}, Z., \& {Shi}, Y. 2011, \mnras, 416, L21,
  \dodoi{10.1111/j.1745-3933.2011.01090.x}

\bibitem[{{White} {et~al.}(2012){White}, {Myers}, {Ross}, {Schlegel},
  {Hennawi}, {Shen}, {McGreer}, {Strauss}, {Bolton}, {Bovy}, {Fan},
  {Miralda-Escude}, {Palanque-Delabrouille}, {Paris}, {Petitjean}, {Schneider},
  {Viel}, {Weinberg}, {Yeche}, {Zehavi}, {Pan}, {Snedden}, {Bizyaev},
  {Brewington}, {Brinkmann}, {Malanushenko}, {Malanushenko}, {Oravetz},
  {Simmons}, {Sheldon}, \& {Weaver}}]{white2012_quasarclustering_boss}
{White}, M., {Myers}, A.~D., {Ross}, N.~P., {et~al.} 2012, \mnras, 424, 933,
  \dodoi{10.1111/j.1365-2966.2012.21251.x}

\bibitem[{{Zhang} {et~al.}(2014){Zhang}, {Gao}, {Henkel}, {Zhao}, {Wang},
  {Menten}, \& {G{\"u}sten}}]{zhang2014_cs_ApJ...784L..31Z}
{Zhang}, Z.-Y., {Gao}, Y., {Henkel}, C., {et~al.} 2014, \apjl, 784, L31,
  \dodoi{10.1088/2041-8205/784/2/L31}

\end{thebibliography}
\bibliographystyle{aasjournal}

\appendix
\section{Signal Attenuation due to Finite Spectral Aperture Size}\label{sec:spectral_attenuation_appendix}

As discussed in Section \ref{sssec:ebossredshifts}, quasars have inherently large velocity offsets between different redshift tracers, due to inflowing and outflowing of different phases of the gas in the galaxy. In the stack, this manifests as an offset of the centroid of the stacked emission in the spectral axis, and a broadening of this emission. This will serve to reduce the signal-to-noise ratio, as emission is spread across spectral channels, and also attenuate the stack luminosity, as these velocity offsets can be wide enough to move signal out of the 7-channel aperture used to calculate the stack luminosity. This effect is compounded by other sources of spectral broadening in the stack, such as the inherent FWHM of the CO line. We defer the full details of the effects of spectral broadening on a stacking analysis to a future work, but we here explore the specific case of eBOSS quasars. 

\begin{figure*}[ht!]
    \centering
    \includegraphics[width=0.8\textwidth]{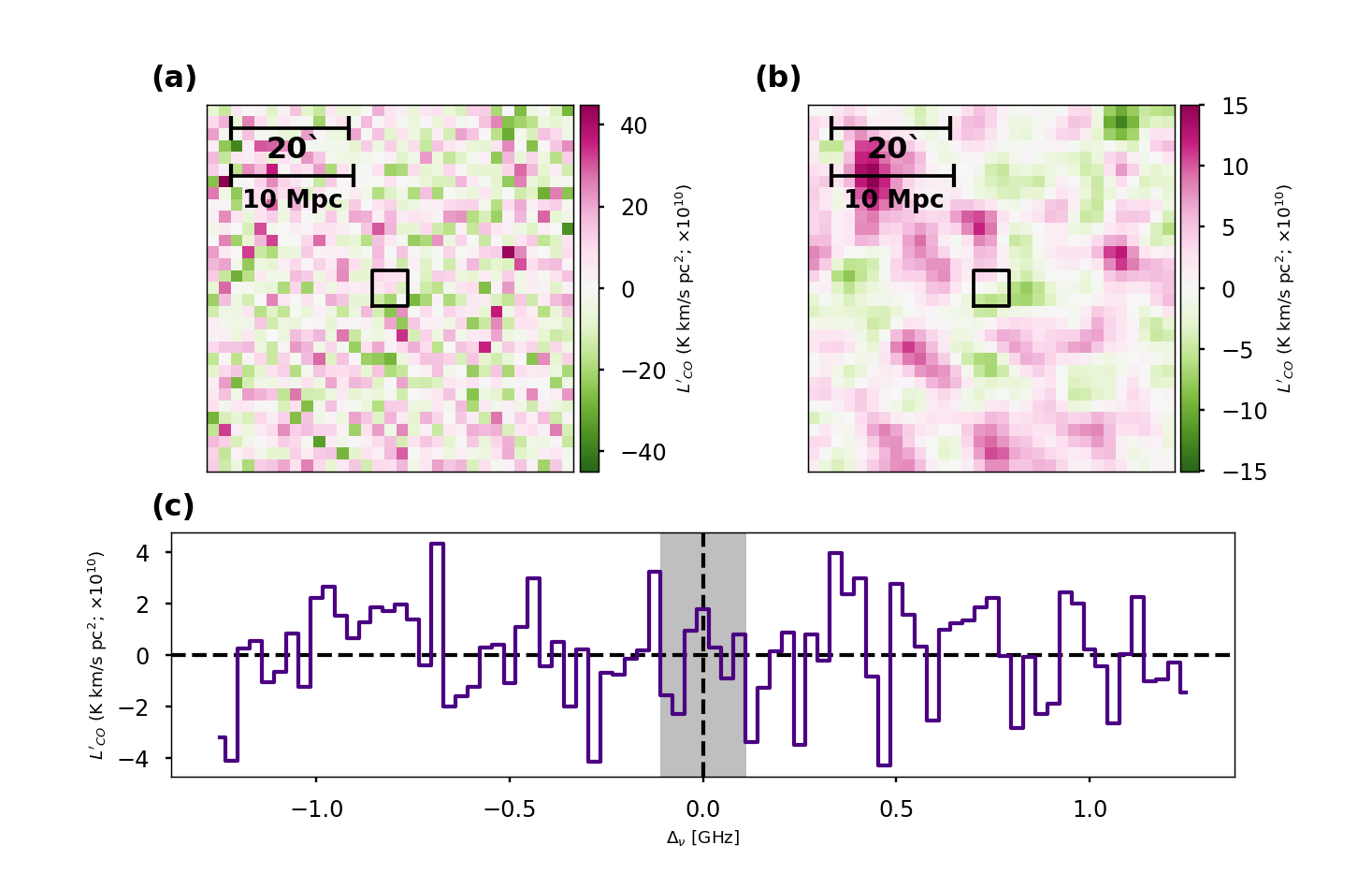}
    \caption{The 3D stack performed on the eBOSS catalog using the $^{12}$CO(1-0) emission wavelength offset by the frequency offset found in individually-detected quasars (see \S\ref{sssec:ebossredshifts}). See Figure \ref{fig:stack_visual} for panel descriptions.}
    \label{fig:offset_stack_visual}
\end{figure*}

\begin{figure*}[ht!]
    \centering
    \includegraphics[width=0.5\textwidth]{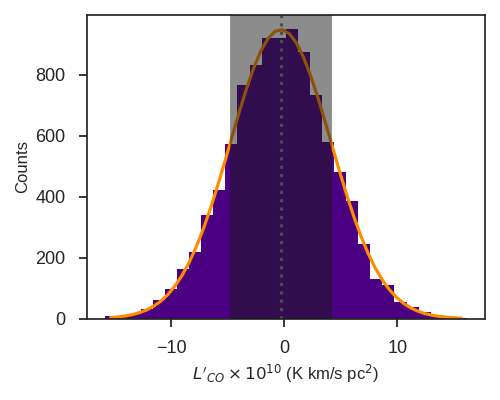}
    \caption{\rev{The bootstrapped uncertainty in \linelum\ for the offset $^{12}$CO(1-0) stack (see Figure \ref{fig:boss_bootstrap_test}).}}
    \label{fig:offset_Bootstrap}
\end{figure*}

We use the previous studies of molecular gas in individual eBOSS quasars shown in Figure \ref{fig:eboss_redshift_comparison} to determine (1) an average CO FWHM for these objects of 319 km/s, (2) an average offset between the molecular redshift and the eBOSS redshift of $\rev{(z_\mathrm{CO}-z_\mathrm{opt})}/(1+z_\mathrm{opt}) = 0.00397$, and (3) a scatter in that offset of $0.00408$. \rev{We note that these correction factors are illustrative values, limited by the small number of eBOSS quasars detected in CO. We would require many more CO measurements of individual quasars to determine a more confident calibration, and these values are not available at the moment. Additionally, we note} that the assembled literature measurements use a variety of tracers which may trace slightly different phases of molecular gas, and thus this scatter may be enhanced compared to a single-tracer study such as COMAP. We deal with the average offset by simply shifting the stack centroid in the spectral axis by this amount. This remains a nondetection (as discussed in Section \ref{sssec:ebossredshifts}); the resulting stack is shown in Figures \ref{fig:offset_stack_visual} and \ref{fig:offset_Bootstrap}. 

\begin{figure*}[ht!]
    \centering
    \includegraphics[width=0.6\textwidth]{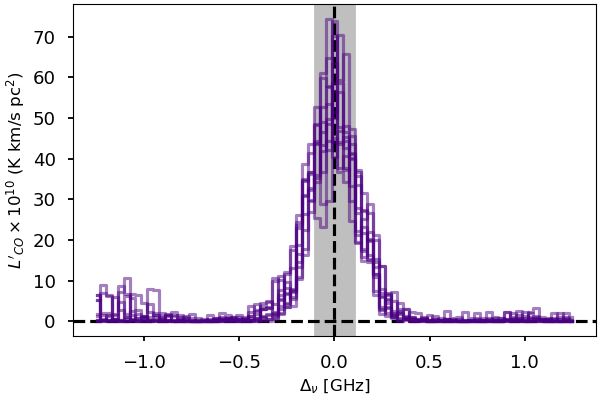}
    \caption{Spectra of simulated stacks, offset to account for the velocity difference between eBOSS-determined redshifts and CO-determined redshifts \rev{$(z_\mathrm{CO}-z_\mathrm{opt})/(1+z_\mathrm{opt}) = 0.00408$)}. Ten different iterations are shown. \rev{The fluctuations in the fringes of the spectrum are due to neighbouring objects falling into some cutouts.}}
    \label{fig:sim_z_offset_spectra}
\end{figure*}

\begin{figure*}[ht!]
    \centering
    \includegraphics[width=0.6\textwidth]{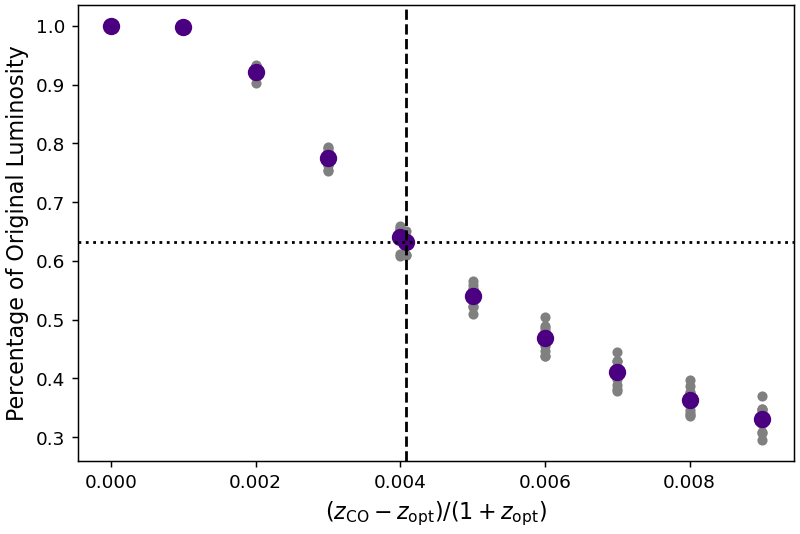}
    \caption{The percentage of the input luminosity returned by a constant-luminosity simulated stack as a function of an induced scatter between $z_\mathrm{opt}$ and $z_\mathrm{CO}$ in the individual halos making up the stack. The average scatter calculated from the independently-detected quasars presented in Section \ref{sssec:ebossredshifts}, taken to be the average of our eBOSS quasar stack, is indicated with the vertical black dashed line. We repeat the test ten times -- the average values are shown as large purple dots.}
    \label{fig:sim_z_offset_testing}
\end{figure*}

To address the scatter in the velocity offset between tracers (functionally, this introduces a redshift uncertainty in the stack), we use the constant luminosity simulations discussed in Section \ref{ssec:simulations} as a framework. We artificially set the linewidth of each CO emitter to be 319 km/s, as measured in Section \ref{sssec:ebossredshifts}. We test the attenuation from various values of scatter in \rev{$(z_\mathrm{CO}-z_\mathrm{opt})/(1+z_\mathrm{opt})$} by offsetting the spectral centroid of each halo indiviudally by a randomly-selected amount pulled from a normal distribution with standard deviation set to the \rev{$(z_\mathrm{CO}-z_\mathrm{opt})/(1+z_\mathrm{opt})$} scatter. This is done before stacking. We then stack the simulated data cube as normal. We perform 10 different iterations of this analysis for each value of \rev{$(z_\mathrm{CO}-z_\mathrm{opt})/(1+z_\mathrm{opt})$} we test, \rev{using the same underlying DM halo distribution as a baseline and varying only the halo velocities}. The resulting spectra for the scatter value we measure from individual eBOSS quasars are shown in Figure \ref{fig:sim_z_offset_spectra}. In Figure \ref{fig:sim_z_offset_testing}, we show how this scatter attenuates signal compared to an un-broadened stack. Using the values determined from individual eBOSS quasars, we find that the returned signal is $63.1\%$ of the actual value, corresponding to an attenuation factor of 1.58.

\section{Cosmic Volume Variation Across the Stack}\label{sec:cosmic_volume_appendix}
One constraint with our current stacking methodology is that the physical volume of each cubelet is redshift-dependent -- the COMAP data cubes stretch from redshift 2.4 to redshift 3.4, so the $6' \times 6' \times  218.75$ MHz aperture we sum to determine quantities such as \linelum\ and \rhoh\ varies in size from $3.0\ \mathrm{Mpc} \times 3.0\ \mathrm{Mpc} \times \rev{2519} \ \mathrm{km/s}$ at low redshifts to $2.7\ \mathrm{Mpc \times 2.7\ Mpc \times 1929\ km/s}$ at high redshifts. We investigate here how this affects the stack. Ideally, this would be addressed by rebinning the data at the map level to create voxels that are constant physical sizes, and we plan to implement this in future work.

As in Appendix \ref{sec:spectral_attenuation_appendix}, we investigate this effect using the constant-luminosity simulations from Section \ref{ssec:simulations}, again artificially imposing a linewidth of 319 km/s for each CO emitter. We generate three different 1000-halo simulations in which emitters are only located in a specific redshift range: a `low-z' simulation with halos $2.39 < z < 2.42$, a `mid-z' simulation with halos $2.82 < z < 2.86$, and a `high-z' simulation with halos $3.89 < z < 3.43$. Each of these redshift ranges corresponds to 12 spectral channels in a COMAP data cube, meaning that the density of sources is the same in each simulated cube. We create three different iterations of each cube at each redshift.

In the spatial axes, the size of the voxel is varying across the stack at a similar rate to the size of the main beam, which is also defined in on-sky coordinates \citep{lamb2021_instrument}. At the COMAP spatial resolution, nearly any astrophysical object is a point source, so the beam defines the spatial distribution of the emission. The distribution of the emission between map voxels is therefore extremely similar across the entire redshift range -- we are stacking together point sources with a similar measured distribution, even if their physical distributions vary. 

\begin{figure*}[ht!]
    \centering
    \includegraphics[width=0.9\textwidth]{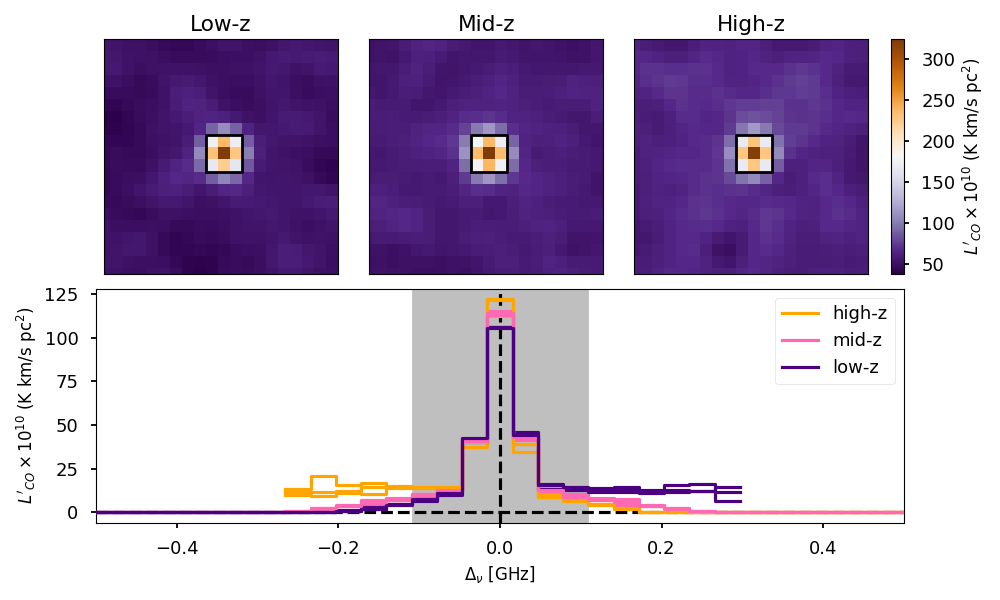}
    \caption{A comparison of constant-luminosity, constant-linewidth simulations at three different redshifts spanning the width of the COMAP data cubes. The spatial distribution of (stacked) emission at each redshift is shown in the top three panels, and the bottom panel shows the spectrum of the stacked cubelets. While the high-redshift stacks are more peaked than the low-redshift stacks, each stack returns the same CO line luminosity. \rev{The low-level fringes in the spectra are background emission from objects other than the one being stacked appearing in the spectrum -- asymmetries appear due to the low-redshift and high-redshift cutoffs. }}
    \label{fig:sim_z_comparison}
\end{figure*}

In the spectral axis, any line broadening is from astrophysical sources and not instrumental sources, as our spectral aliasing is near-zero. At low redshifts, this is $1.15\times$ the width of a map channel, whereas at high redshifts it is $0.9\times$ the width of a channel. An identical CO emitter will thus (unphysically) appear more strongly peaked at high redshifts than at low redshifts (Figure \ref{fig:sim_z_comparison}). However, the same amount of emission is still present (spread between more spectral channels) and our 7-channel aperture is more than enough to encompass this spread, so the returned \linelum\ value remains the same.

While the line luminosity is unaffected by the change in cosmic volume across the stack, the resulting molecular gas density, \rhoh, does vary across the stack, because the volume being averaged over varies across the stack. This is because, in our simulations, all of the input luminosity is coming from a single source, but being spread over variable volumes across the stack due to the main beam. Volumes vary by roughly 37\% across the full range of the stack -- the same percentage difference we observe in the \rhoh\ values output from these simulations.

\section{Foreground Stacks}\label{app:foregroundstacks}
As discussed in Section \ref{ssec:foregrounds}, there are several other molecular spectral lines whose emission may fall into the COMAP frequency range. These spectral lines could potentially contaminate the expected CO(1-0) LIM signal. Conversely, these spectral lines are important tracers of galaxy properties in their own rights, and any signal would be interesting and informative. We present stacks on each of these foregrounds lines below.

\newpage
\subsection{HCN(1-0)}\label{assec:hcn_foregrounds}

HCN(1-0) is a tracer of dense ($n(\mathrm{H_2}) \geq 3\times 10^4 \ \mathrm{cm^{-3}}$) molecular gas in galaxies, associated primarily with the star-forming cores of Giant Molecular Clouds \citep{gaosolomon2004_hcn}. HCN luminosity correlates strongly with CO luminosity, although in a given galaxy $L'_\mathrm{HCN}$ is roughly an order of magnitude less than \linelum. HCN has a rest frequency of 88.63 GHz, and 595 eBOSS quasars with this rest frequency fall into the COMAP volume. 

\begin{figure*}[ht!]
    \centering
    \includegraphics[width=0.8\textwidth]{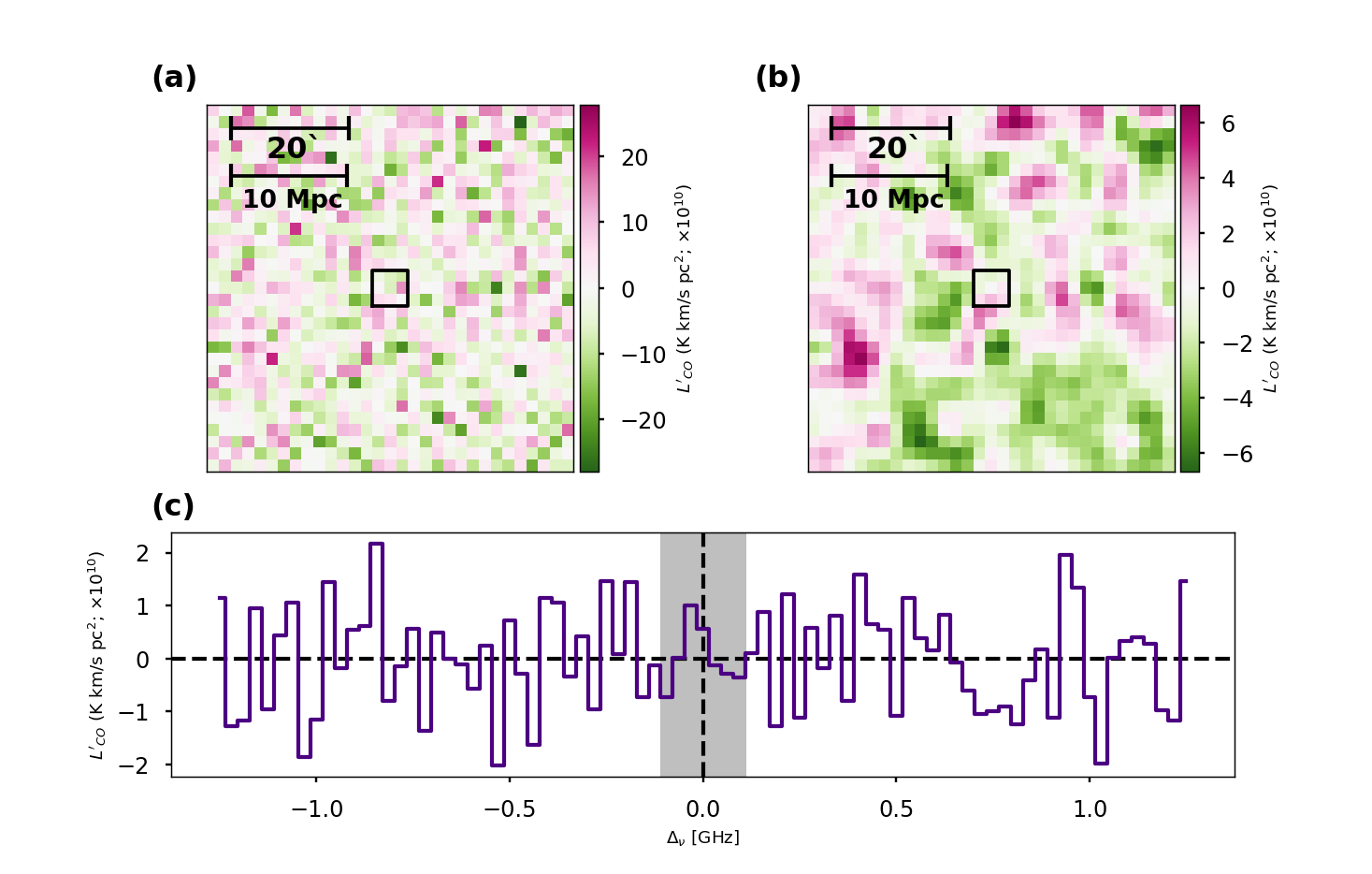}
    \caption{The 3D stack performed on the eBOSS catalog using the HCN(1-0) emission wavelength. See Figure \ref{fig:stack_visual} for panel descriptions.}
    \label{fig:hcn_stack_visual}
\end{figure*}

\begin{figure*}[ht!]
    \centering
    \includegraphics[width=0.5\textwidth]{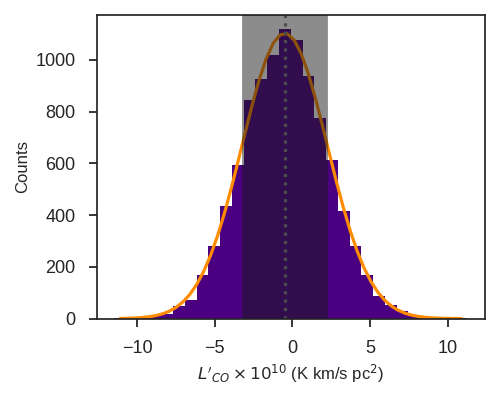}
    \caption{\rev{The bootstrapped uncertainty in \linelum\ for the HCN(1-0) stack (see Figure \ref{fig:boss_bootstrap_test}).}}
    \label{fig:hcn_bootstrap}
\end{figure*}

\newpage
\subsection{CS(2-1)}\label{assec:cs_foregrounds}
CS is interesting as a dense gas tracer independent of HCN/CN$^-$ chemistry, although it is quite faint \citep[e.g.]{wang2011_cs_MNRAS.416L..21W, zhang2014_cs_ApJ...784L..31Z}). With a rest frequency of of 91.98 GHz, there are 479 eBOSS objects included in this stack.

\begin{figure*}[ht!]
    \centering
    \includegraphics[width=0.8\textwidth]{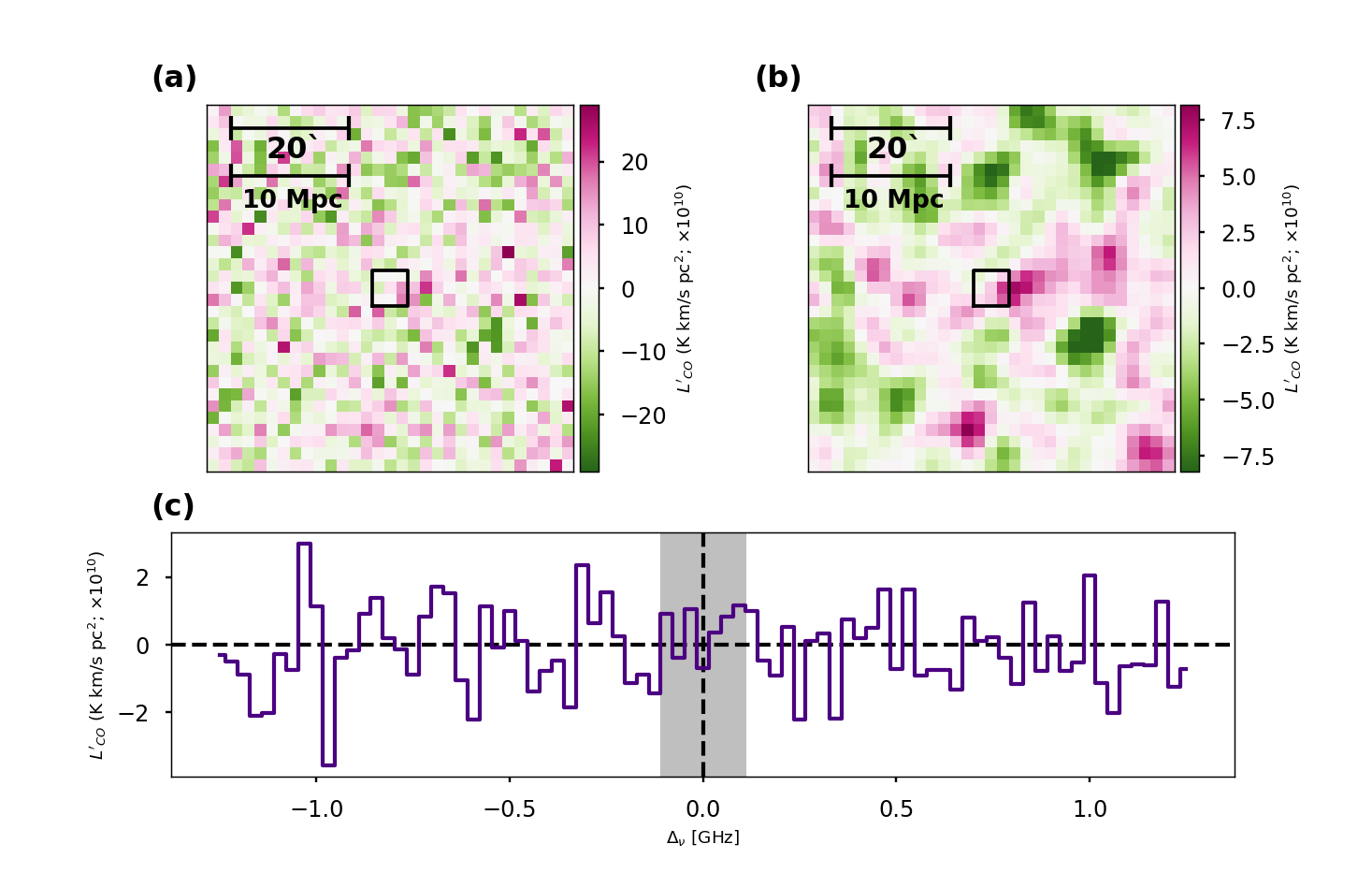}
    \caption{The 3D stack performed on the eBOSS catalog using the CS(2-1) emission wavelength. See Figure \ref{fig:stack_visual} for panel descriptions.}
    \label{fig:cs_stack_visual}
\end{figure*}

\begin{figure*}[ht!]
    \centering
    \includegraphics[width=0.5\textwidth]{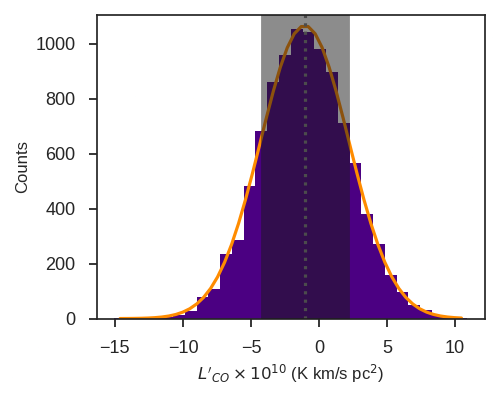}
    \caption{\rev{The bootstrapped uncertainty in \linelum\ for the CS(2-1) stack (see Figure \ref{fig:boss_bootstrap_test}).}}
    \label{fig:cs_bootstrap}
\end{figure*}

\newpage
\subsection{$^{13}$CO(1-0)}\label{assec:13co_foregrounds}

\thirteenco, as a rarer and slightly heavier isotopologue of $^{12}$CO, has a similar rest frequency (110.20 GHz) and traces a similar phase of gas, although with a higher critical density. As an isotopologue, \thirteenco\ is of particular interest in determining the CO-H$_2$ conversion factor in galaxies \citep[e.g.][]{israel2020_13coratio_A&A...635A.131I}. 311 eBOSS objects are included in this stack.

\begin{figure*}[ht!]
    \centering
    \includegraphics[width=0.8\textwidth]{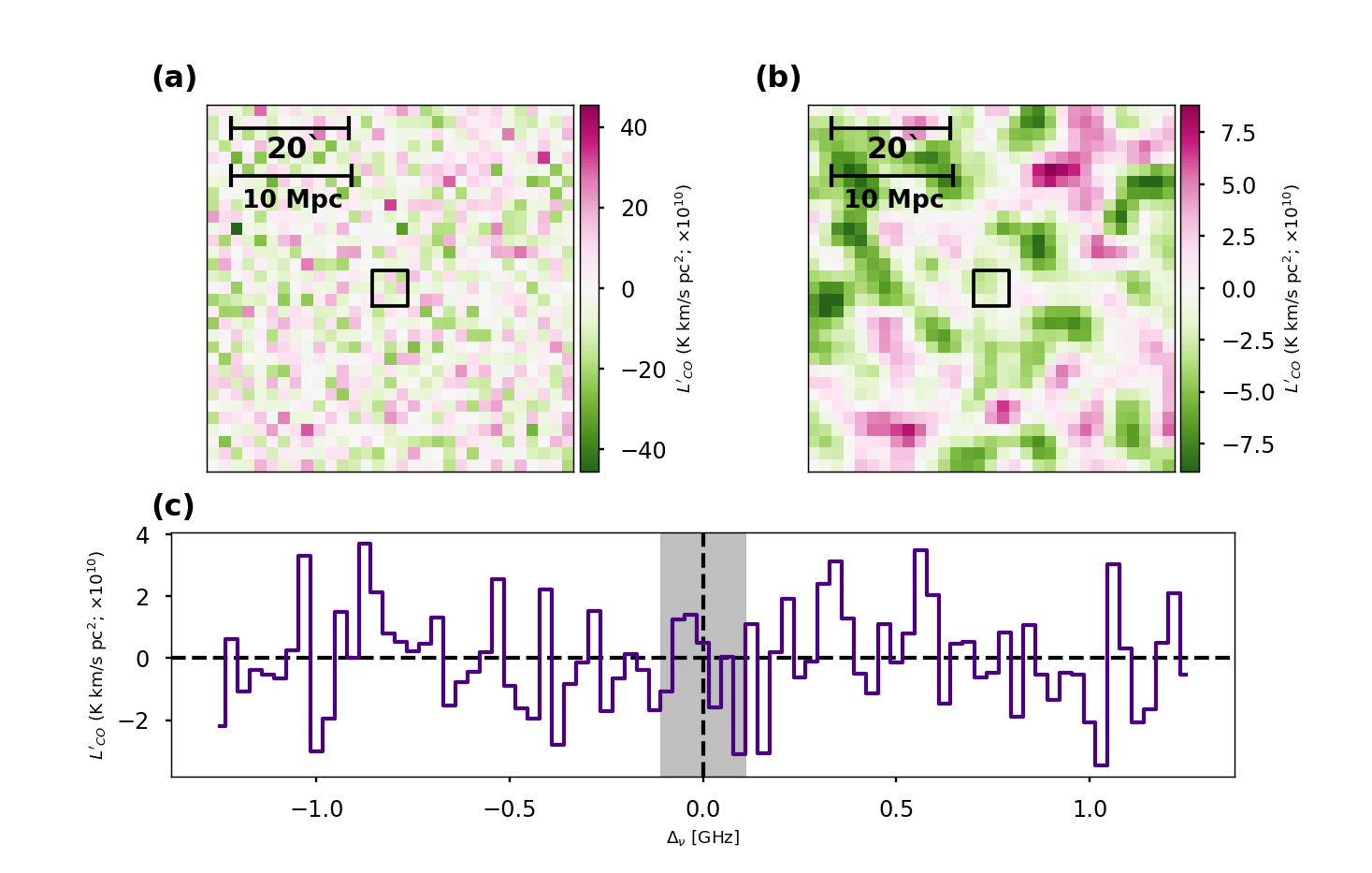}
    \caption{The 3D stack performed on the eBOSS catalog using the $^{13}$CO(1-0) emission wavelength. See Figure \ref{fig:stack_visual} for panel descriptions.}
    \label{fig:13co_stack_visual}
\end{figure*}

\begin{figure*}[ht!]
    \centering
    \includegraphics[width=0.5\textwidth]{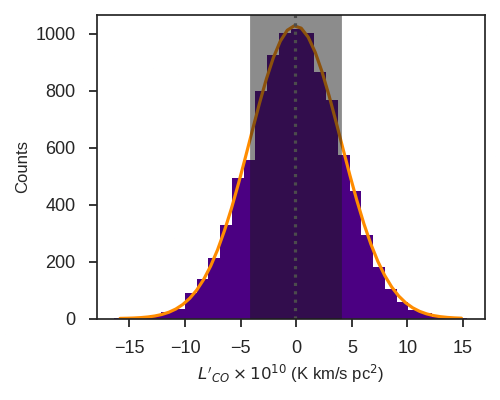}
    \caption{\rev{The bootstrapped uncertainty in \linelum\ for the $^{13}$CO(1-0) stack (see Figure \ref{fig:boss_bootstrap_test}).}}
    \label{fig:13co_bootstrap}
\end{figure*}

\newpage
\subsection{CN$^-$(1-0)}\label{assec:cn_foregrounds}

CN$^-$(1-0) traces a very similar phase of gas to HCN (i.e., dense molecular cores; \S\ref{assec:hcn_foregrounds}), but has a critical density which is lower by roughly a factor of 5, and likely additionally prefers gas layers affected by stellar UV radiation \citep{riechers2007_cn_ApJ...666..778R}. The CN$^-$ rest frequency of 113.49 GHz is very close to CO(1-0), and 269 eBOSS quasars fall into the COMAP volume at this frequency.

\begin{figure*}[ht!]
    \centering
    \includegraphics[width=0.8\textwidth]{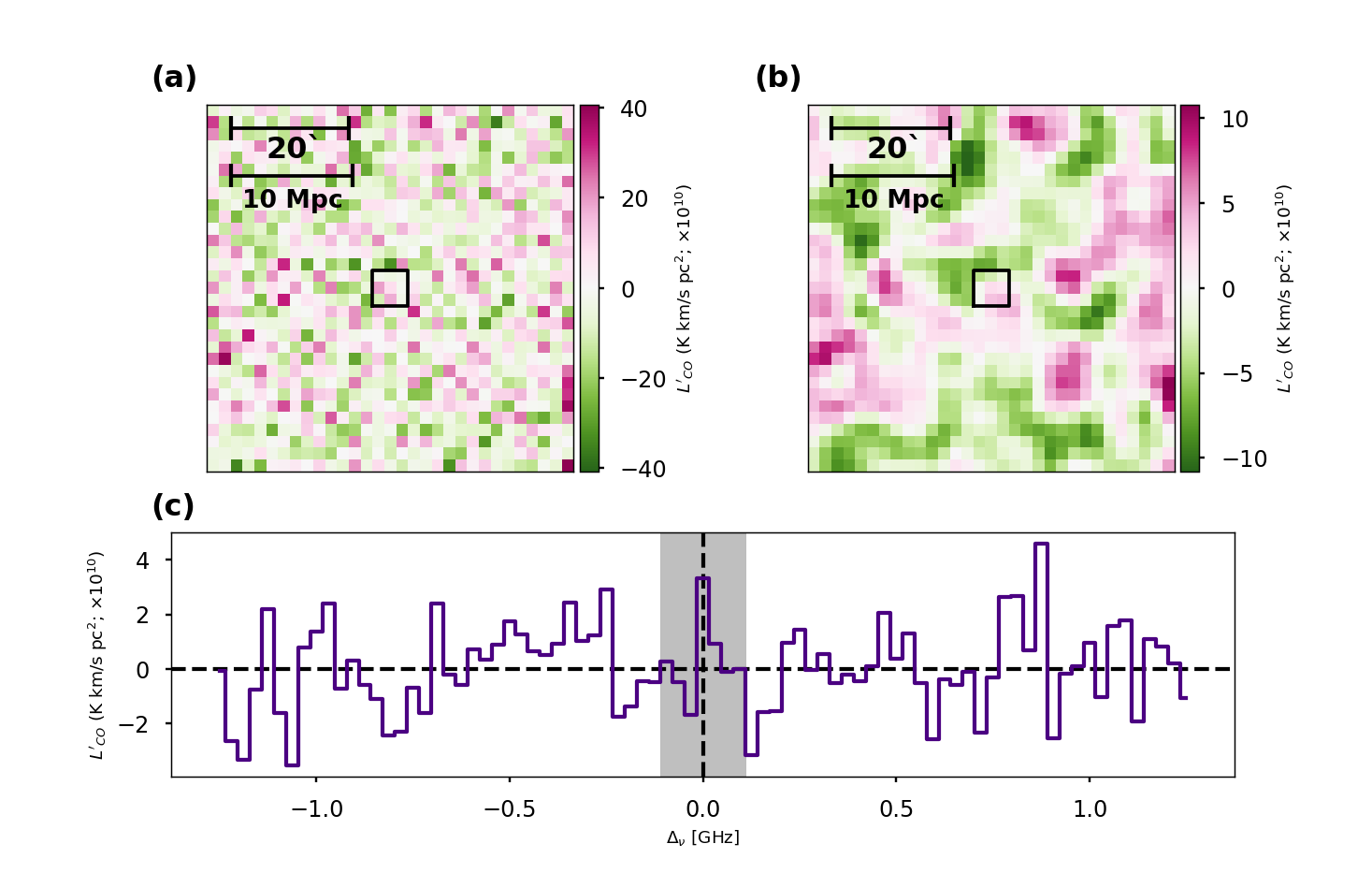}
    \caption{The 3D stack performed on the eBOSS catalog using the CN$^-$(1-0) emission wavelength. See Figure \ref{fig:stack_visual} for panel descriptions.}
    \label{fig:cn_stack_visual}
\end{figure*}

\begin{figure*}[ht!]
    \centering
    \includegraphics[width=0.5\textwidth]{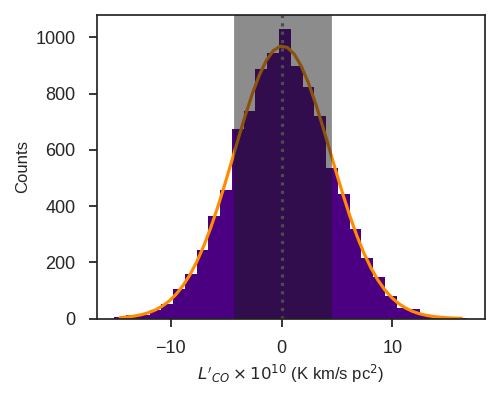}
    \caption{\rev{The bootstrapped uncertainty in \linelum\ for the CN$^-$(1-0) stack (see Figure \ref{fig:boss_bootstrap_test}).}}
    \label{fig:cn_bootstrap}
\end{figure*}

\end{document}